\documentclass[a4paper, twocolumn, 11pt, accepted=2022-11-16]{quantumarticle}
\pdfoutput=1
\usepackage[numbers, sort&compress]{natbib}
\usepackage[utf8]{inputenc}
\usepackage[english]{babel}
\usepackage[T1]{fontenc}
\usepackage{amsmath,amssymb,amsthm}
\usepackage{hyperref}
\usepackage{dsfont}
\usepackage{mathrsfs}
\usepackage{enumitem}

\usepackage[acronym]{glossaries}
\glsdisablehyper
\newacronym{spam}{SPAM}{State Preparation and Measurement}
\newacronym{rb}{RB}{Randomized Benchmarking}
\newacronym{asf}{ASF}{Average Sequence Fidelity}
\newacronym{cp}{CP}{Completely Positive}
\newacronym{cptp}{CPTP}{Completely Positive Trace Preserving}
\newacronym{povm}{POVM}{Positive Operator Valued Measurement}
\newacronym{mpo}{MPO}{Matrix Product Operator}

\newcommand*{\fid}{\boldsymbol{\mathsf{F}}}
\newcommand*{\f}{\frac}
\newcommand*{\mc}{\mathcal}
\newcommand*{\dg}{\dagger}

\newcommand*{\mbb}{\mathds}
\newcommand*{\mscr}{\mathscr}
\newcommand*{\ex}{\mathrm{e}}
\newcommand*{\dimE}{d_\mathsf{E}}
\newcommand*{\dimS}{d_\mathsf{S}}
\newcommand*{\env}{\mathsf{E}}
\newcommand*{\syst}{\mathsf{S}}
\DeclareMathOperator{\tr}{tr}
\DeclareMathOperator*{\avg}{\mbb{E}}
\DeclareMathOperator*{\Motimes}{\text{\raisebox{0.25ex}{\scalebox{0.6}{$\bigotimes$}}}}
\DeclareMathOperator*{\Moplus}{\text{\raisebox{0.25ex}{\scalebox{0.6}{$\bigoplus$}}}}
\DeclareMathOperator*{\Mcirc}{\text{\raisebox{0.25ex}{\scalebox{0.7}{$\bigcirc$}}}}
\DeclareMathOperator*{\llangle}{\langle\!\langle\!}
\DeclareMathOperator*{\rrangle}{\!\rangle\!\rangle}

\usepackage{accents}
\newcommand\undervec[1]{\underaccent{\vec}{#1}}
\newcommand*{\markov}{{\scriptscriptstyle{(\mathrm{Mkv})}}}

\newtheorem*{theorem}{Theorem}
\newtheorem*{definition}{Definition}
\newtheorem*{corollary}{Corollary}
\newtheorem*{lemma}{Lemma}
\newtheorem*{proposition}{Proposition}
\newtheorem{notation}{Notation}

\begin{document}

\title{Towards a general framework of Randomized Benchmarking incorporating non-Markovian Noise}

\author{Pedro Figueroa-Romero}
\affiliation{Hon Hai Quantum Computing Research Center, Taipei, Taiwan}
\orcid{0000-0002-1386-4380}
\thanks{Current address - IQM Germany GmbH, Nymphenburgerstrasse 86, 80636 Munich, Germany. \href{malto:pedro.romero@meetiqm.com}{pedro.romero@meetiqm.com}}
\author{K. Modi}
\email{kavan.modi@monash.edu}
\affiliation{School of Physics and Astronomy, Monash University, Clayton, VIC 3800, Australia}
\affiliation{Centre for Quantum Technology, Transport for New South Wales, Sydney, NSW 2000, Australia}
\orcid{0000-0002-2054-9901}
\author{Min-Hsiu Hsieh}
\affiliation{Hon Hai Quantum Computing Research Center, Taipei, Taiwan}
\email{min-hsiu.hsieh@foxconn.com}
\orcid{0000-0002-3396-8427}
\maketitle
\begin{abstract}
The rapid progress in the development of quantum devices is in large part due to the availability of a wide range of characterization techniques allowing to probe, test and adjust them. Nevertheless, these methods often make use of approximations that hold in rather simplistic circumstances. In particular, assuming that error mechanisms stay constant in time and have no dependence in the past, is something that will be impossible to do as quantum processors continue scaling up in depth and size. 
We establish a theoretical framework for the \gls{rb} protocol encompassing temporally-correlated, so-called non-Markovian noise, at the gate level, for any gate set belonging to a wide class of finite groups. We obtain a general expression for the \gls{asf} and propose a way to obtain average fidelities of full non-Markovian noise processes. 
Moreover, we obtain conditions that are fulfilled when an \gls{asf} displays authentic non-Markovian deviations. Finally, we show that even though gate-dependence does not translate into a perturbative term within the \gls{asf}, as in the Markovian case, the non-Markovian sequence fidelity nevertheless remains stable under small gate-dependent perturbations.
\end{abstract}
\glsresetall

\section{Introduction}
Having effective and efficient methods for characterization, verification and validation, is of vital importance for the production of useful quantum devices, and it will remain an integral part of them, serving as their debugging component. The \gls{rb} protocol~\cite{Emerson_2005, Levi_2007, Knill2008, PhysRevLett.106.180504, PhysRevA.85.042311} has become the method par-excellence to begin a characterization process, being a highly economical method in many respects, in comparison e.g., with Quantum Process Tomography~\cite{Chuang_1997} or Gate Set Tomography~\cite{nielsen2020gate}, despite providing only a limited amount information. In essence, \gls{rb} consists of a series of steps whereby sequences of gates drawn at random are applied and averaged over many sequence runs, with the outputs allowing to estimate average error rates within the device. The simplicity of the core idea of \gls{rb} led to a plethora of methods for different specific purposes, for different sets of gates or under different noise profiles~\cite{helsen2020general}.

A remarkable feature of standard \gls{rb} under simplifying assumptions is that the profile of the data produced by it, which we call an \gls{asf}, follows an exponential decay in the number of gates that are applied. Furthermore, in this case, \gls{spam} errors do not affect such decay~\cite{Emerson_2005, Levi_2007, Knill2008, PhysRevLett.106.180504, PhysRevA.85.042311}. This is not a universal feature of \gls{rb}, but rather it is a consequence of both the particular gates being benchmarked and the approximations made regarding the underlying noise. In the simplest case, these correspond to multi-qubit Clifford gates, and assumptions such as time-independence, gate-independence and Markovianity\footnote{Noise not depending on \emph{when} or on \emph{what} gates were applied, and noise \emph{forgetting} previous noise, respectively.}. Beyond these assumptions, it is known that within the Markovian approximation, the decays stay relatively benign, whereby time-dependent \gls{asf}s become an exponential with changing decay rates between steps, and gate-dependence simply adds up a perturbative term, itself decaying exponentially in sequence length.

Most literature to date on noise characterization techniques, including \gls{rb}, begin by assuming Markovianity. The characterization of errors beyond this has largely remained unexplored. Due to the very fragility of open quantum systems and the often unstable nature of their surroundings, the Markovian assumption can quickly become ineffective, rendering standard methods unreliable. Indeed, time dependence in effects such as drift~\cite{van_Enk_2013, fogarty2015nonexponential, Proctor2020}, crosstalk~\cite{PhysRevLett.109.240504, Sarovar2020detectingcrosstalk, ParradoRodriguez2021crosstalk}, leakage~\cite{PhysRevA.97.032306, Wallman_leakage, PhysRevA.92.042333}, as well as other memory effects, is an ubiquitous and unavoidable feature~\cite{osti_1671379, Ryan_2009, Bylander2011, fogarty2015nonexponential, PhysRevB.92.035442, PhysRevApplied.10.044017, PhysRevB.97.180505, Burnett2019, Proctor2020} of open quantum systems. Therefore, a framework for the characterization of temporally-correlated (non-Markovian) errors becomes a crucial aspect to address in order to reach scalability and reliability in quantum devices.

In this regime of non-Markovian noise, both particular cases~\cite{fong2017randomized, Mavadia_2018, PhysRevA.93.022303, PhysRevA.103.022607}, as well as the general \gls{asf}~\cite{figueroaromero2021randomised}, have been studied, showing that the simplicity of \gls{rb} can still be exploited. These, however, are, to date, restricted to the gate sets forming a unitary 2-design, of which the multi-qubit Clifford group is an example~\cite{graydon2021clifford}. Furthermore, as in the case for multi-qubit Cliffords, these gates are often built-up from other smaller generators, rather than being themselves simpler operations. While a generalization to gate sets forming finite groups in the case of Markovian noise has been addressed in quite some generality~\cite{hashagen_finite, Helsen2019, helsen2020general}, it has been hitherto unclear what happens in the non-Markovian scenario, and whether this regime can also be incorporated into a more general \gls{rb} framework.

In this manuscript, we establish a framework for \gls{rb} under generic non-Markovian noise, as long as it is gate-independent, for any gate set that forms a finite group admitting a multiplicity-free representation. This framework generalizes both the Markovian and non-Markovian cases for standard gate-independent \gls{rb} considered thus far, and allows to tailor for most specific-purpose \gls{rb} protocols developed within the Markovian assumption~\cite{helsen2020general}. In particular, it lets us swiftly obtain a general \gls{asf}, propose an operational non-Markovian fidelity figure of merit, and obtain conditions under which noise will produce uniquely non-Markovian \gls{rb} data. Furthermore, given that gate-dependence and more general context-dependence are also generally unavoidable in realistic scenarios, we argue that the Markovian, gate-dependent, results of~\cite{Wallman_2018, PhysRevLett.119.130502} do not generalize trivially to the non-Markovian case. Nevertheless, we show stability under small gate-dependent perturbations, and discuss potential ways for general gate-dependence to be incorporated within our framework.

We thus establish a comprehensive, albeit by no means exhaustive, theoretical framework for \gls{rb} with a wide class of gate sets for temporally-correlated noise at the gate level. Given the pressing need to understand time-dependent noise effects in quantum technologies, as these scale-up in size and depth, this seeks to eventually incorporate non-Markovianity within a general framework for \gls{rb}, in the spirit of~\cite{helsen2020general}, or further within a more general framework for quantum device characterization.

We begin with a review of standard \gls{rb} and non-Markovianity in Section~\ref{sec: rb and nM review}. We then present our main results, which are distributed as follows:

\begin{itemize}[label={$\diamond$},leftmargin=1em,labelwidth=*,align=left]
    \item In Section~\ref{sec: asf}, we obtain a non-Markovian and gate-independent \gls{asf} for finite groups which clearly captures the role of the environment in carrying information about the average noise in a \gls{rb} experiment. We do this by introducing quantities called \emph{quality maps}, the central objects carrying average noise rates and the temporal correlations therein.
    
    \item In Section~\ref{sec: avg gate fidelity}, we propose a method to operationally quantify average non-Markovian gate fidelities of full \gls{rb} sequences by consistently averaging over initial states and measurements.
    
    \item In Section~\ref{sec: nM}, we  obtain sufficient and necessary conditions on the noise in order to witness authentically non-Markovian deviations in the \gls{asf}. In particular, we find that the noise on the full environment plus system has to be such that intermediate states increase in purity by at least the square of the size of the system.
    
    \item In Section~\ref{sec: gate-dependence}, we incorporate gate-dependence and argue that the results in~\cite{PhysRevLett.119.130502, Wallman_2018}, showing that gate-dependence induces a single exponentially vanishing perturbative term on a Markovian \gls{asf}, do not carry on to the realm of non-Markovian noise. Nevertheless, we show that both the \gls{asf} and the variance of the sequence fidelity remain stable under small gate-dependent perturbations in the noise.
\end{itemize}

Finally, in Section~\ref{sec: conclusions}, we give some perspective of our results and propose ways of moving forward in the treatment of non-Markovian noise within characterization protocols and techniques.

\section{An overview of Randomized Benchmarking and non-Markovianity}\label{sec: rb and nM review}

\subsection{Standard RB}
Consider a sequence of $m$ quantum gates, $\mc{G}_m:=\Mcirc_{i=1}^m \mc{C}_i:=\mc{C}_m\circ\cdots\circ\mc{C}_1$ with $\circ$ denoting gate composition, followed by an inverse sequence, i.e., an undo-gate, $\mc{C}_{m+1}:=\Mcirc_{i=m}^1\mc{C}_i^{-1}$. This amounts to an overall map that is the identity gate. In the standard gate-independent \gls{rb} protocol (shown in detail in Appendix~\ref{appendix: rb protocol}), one considers a composition sequence $\mc{S}_m := \Mcirc_{i=1}^{m+1}\left(\Lambda_i\circ\mc{C}_i\right)$, where the \gls{cptp} maps $\Lambda_i$ model gate-independent noise inherent to the physical realization of the gates. This is equivalently expressed as $\mc{S}_m=\Lambda_{m+1}\Mcirc_{i=1}^m(\mc{G}_i^\dg\circ\Lambda_i\circ\mc{G}_i)$. Experimentally, for a suitable initial state $\rho$ and a \gls{povm} element $M$, the protocol outputs give an estimate of the \gls{asf},
\begin{equation}
    \mc{F}_m:=\tr\{M\mbb{E}[\mc{S}_m(\rho)]\},
    \label{eq: asf}
\end{equation}
where averaging, denoted by $\mbb{E}$, is taken over all gates $\mc{G}_i$. This is the central quantity in standard \gls{rb}.

Whenever the gates $\mc{G}_i$ belong to a unitary 2-design, i.e., a set with identical second moments as the uniform unitary group, and when the noise is time-independent, i.e., $\Lambda_1 = \Lambda_2 = \cdots=\Lambda_{m+1}$, the {\gls{asf}} takes the form of a decaying exponential in the sequence length, with the rate of decay capturing the average gate fidelity of the physical gates with respect to the ideal ones\footnote{ There are some caveats to this statement having to do with the concept of gauge-freedom in the representation of the gates; for detail, see e.g., the section ``Randomized Benchmarking and Average Fidelity'' of~\cite{helsen2020general}.}, and the \gls{spam}~errors being absorbed in both a multiplicative and an offset constants. In general, the functional form of the {\gls{asf}} depends not only on the specific gate set to be benchmarked, but also on the assumptions made about the noise. Both a class of non-Clifford gate sets has been considered~\cite{ScalableNonClif, PhysRevA.97.062323, Hashagen_2018,  hashagen_finite,  PhysRevA.92.060302, PhysRevLett.109.240504, Helsen2019, helsen2020general} and the assumptions on the noise relaxed, e.g., for time-dependent~\cite{PhysRevA.85.042311,PhysRevLett.106.180504,Wallman_2014}, gate-dependent~\cite{PhysRevLett.119.130502,Wallman_2018,PhysRevLett.106.180504, merkel2018randomized, Carignan_Dugas_2018} or non-Markovian noise~\cite{PhysRevA.103.022607,PhysRevA.93.022303,figueroaromero2021randomised}, although to this day, arguably the least explored regime is that of non-Markovian noise.

\subsection{Non-Markovian Quantum Processes}

Non-Markovianity generally refers to a dependence of subsequent outcomes on previous ones, and in the context of {\gls{rb}}, it implies that the noise at a given step is temporally correlated with the noise that preceded it and that the data outputs cannot be obtained by modeling noise as local quantum channels. The functional form of the {\gls{asf}} for unitary 2-designs in the non-Markovian regime is not that of a decaying exponential in sequence length anymore, but rather that of a non-trivial function of the memory within the noise~\cite{figueroaromero2021randomised}: this makes the extraction of operationally meaningful figures of merit a more elaborated task than in the Markovian case, despite the simplicity of the {\gls{rb}} protocol. Nevertheless, the mere fact that no physical system can be completely isolated from its surroundings requires considering the presence of temporal correlations.

Classically, non-Markovianity can be described by a stochastic process $\{X_t\}$ where information is being sent between timesteps such that the state of the system is conditionally dependent on the past, i.e.
\begin{equation}
    \mbb{P}(x_k|x_{k-1},\ldots,x_0) = \mbb{P}(x_k|x_{k-1},\ldots,x_{k-\ell}),
    \label{eq: classical nM definition}
\end{equation}
for any integers $0\leq{\ell}\leq{k}$ and sequences of event outcomes $x_i$, with $\mbb{P}(\cdot|\cdot)$ denoting a conditional probability. In particular, when $\ell=1$, the process is called \emph{Markovian} and when $\ell=0$ it is called \emph{random}; otherwise, the process is non-Markovian with Markov order $\ell$. The fact that the complexity in describing non-Markovian processes increases exponentially in increasing Markov-order can be seen from joint probabilities requiring up to $\ell$-point correlations within the respective conditional probabilities.

Quantum mechanically, the process tensor\footnote{The object we call process tensor is also known in different settings as quantum comb~\cite{PhysRevLett.101.060401, PhysRevA.80.022339}, causal box~\cite{Portmann_causal}, correlation kernel~\cite{nurdin2021heisenberg}, process matrix~\cite{Costa_2016}, channel with memory~\cite{PhysRevA.72.062323}, or strategy~\cite{Gutoski_games}, to mention some.} framework~\cite{PhysRevA.97.012127, PhysRevLett.120.040405, Milz_2017, Milz2020kolmogorovextension, taranto2019memory, figueroaromero2021equilibration, milz2020quantum, PhysRevLett.123.040401}, takes into account the invasive nature of observation to unambiguously provide a generalization of the condition in Eq.~\eqref{eq: classical nM definition}, as shown in~\cite{PhysRevLett.120.040405}. In this case, the medium for information to be sent across timesteps is an environment $\env$, defined by a Hilbert space $\mscr{H}_\env$ part of a bipartite closed system $\mscr{H}_\env\otimes\mscr{H}_\syst$, with $\syst$ being the system of interest. Henceforth we set the respective dimensions as $\dimE\dimS:=\dim(\mscr{H}_\env\otimes\mscr{H}_\syst)$. Then, for an initial state $\rho$ of $\syst\env$, and upon measuring a {\gls{povm}} $\mc{J}_k:=\{M_{x_n}^{(k)}\}_{x_n}$ on system $\syst$, we may describe the probability of observing a sequence of quantum events $x_k,\ldots,x_0$ by
\begin{equation}
    \mbb{P}(x_k,\ldots,x_0|\mc{J}_k,\ldots,\mc{J}_0) := \tr\left[M_{x_k}\rho^{(k)}\right],
    \label{eq: joint prob quantum}
\end{equation}
where $\rho^{(k)}:=\tr_\env[\Mcirc_{i=1}^{k-1}(\mc{U}_i\circ\mc{A}_{x_i})\rho]$ is the state of system $\syst$ at the $k$\textsuperscript{th} timestep, with $\mc{U}_i$ being unitary maps on $\syst\env$ describing the evolution of the full system between timesteps and $\mc{A}_{x_i}$ being \gls{cp} maps acting on system $\syst$ alone: precisely, each $\mc{J}_i:=\{\mc{A}^{(i)}_{x_n}\}_{x_n}$ is called an instrument, where $\mc{A}^{(i)}_{x_n}$ is an experimental intervention represented by a \gls{cp} map with state outcome $x_n$, and such that $\sum_{x_n}\mc{A}^{(i)}_{x_n}=\mc{A}^{(i)}$ is a \gls{cptp} map.

We drop the super-indices in Eq.~\eqref{eq: joint prob quantum} for clarity, which we may write more succinctly  as the inner product
\begin{align}
    \mbb{P}(x_k,\ldots,x_0|\mc{J}_k,\ldots,\mc{J}_0) = \tr\left(\Upsilon_k\,\Theta_k^\mathrm{T}\right),
    \label{eq: probabilities PT contraction}
\end{align}
where $\mathrm{T}$ denotes a transpose, and $\Upsilon_k$ and $\Theta_k$ are tensors containing all dynamics $\{\mc{U}_i\}$ and all interventions $\{\mc{A}_i\}$; in the Choi-Jamio\l{}kowski representation, these take the form
\begin{align}
    \Upsilon_k &:= \tr_\env\left\{\left[\Mcirc_{i=1}^k(\mc{U}_i\otimes\mc{I}_{\mathsf{aux}}\circ\mscr{S}_i)\right]\rho\otimes\psi^{\otimes{k}}\right\},
    \label{eq: Choi state pt}\\
    \Theta_k &:= M_{x_k} \otimes \left[\Motimes_{i=1}^{k-1}\left(\mbb1_{\mathsf{A}_i}\otimes\mc{A}_{x_i}\right)\right]\psi^{\otimes{k}},
\end{align}
where $\mathsf{aux}:=\mathsf{A}_1\mathsf{B}_1\ldots\mathsf{A}_k\mathsf{B}_k$, with $\mathsf{A}_i$, $\mathsf{B}_i$ being $\dimS$-dimensional auxiliary spaces, $\mscr{S}_i$ being a swap map between $\syst$ and $\mathsf{A}_i$, and $\psi=\sum|ii\rangle\!\langle{jj}|$ being an unnormalized maximally entangled state.

The process tensor framework thus allows to neatly separate the underlying dynamical source for any given quantum process, including all temporal correlations therein, from all experimentally controllable operations. This description is entirely general as a quantum stochastic process framework~\cite{Milz2020kolmogorovextension, milz2020quantum}, and similarly the instruments used to describe interventions are entirely general and can be temporally correlated themselves. Similar to the case of quantum states, the choice of employing a Choi state representation in Eq.~\eqref{eq: Choi state pt} allows us to readily deduce properties of the process. In particular, temporal properties get codified as spatial properties within the Choi state, so that a Markov process takes an uncorrelated form, $\Upsilon^\markov:=\Motimes_i\mscr{Y}_{i:i-1}\otimes\rho_\syst$, with $\mscr{Y}_{i:i-1}$ being individual Choi states of dynamics connecting the $(i-1)$\textsuperscript{th} and $i$\textsuperscript{th} steps. This implies that we may quantify the non-Markovianity of a process by simply quantifying its distinguishability from the closest Markovian one, i.e., $\mc{N}:=\min_{\Upsilon^\markov}d(\Upsilon,\Upsilon^\markov)$, for any operationally meaningful distance measure $d(\cdot,\cdot)$.

\subsection{Non-Markovian RB}

As stochastic processes are ubiquitous in science, the process tensor framework has proven useful in a wide range of topics, from the foundational~\cite{Milz2020kolmogorovextension, 2019almostmarkovian, PhysRevE.102.032144, milz2020genuine, PhysRevX.10.041049, milz2021resource}, to the applied in the characterization and control of quantum devices~\cite{White_2020, PhysRevA.102.062414, white2021nonmarkovian, berk2021extracting}. In the case of \gls{rb}, it is clear that we can describe the \gls{asf} in Eq.~\eqref{eq: asf} as a contraction of process tensors without a need to assume Markovianity for the noise. That is, we now have
\begin{equation}
    \mc{S}_m \!=\! \tr_\env\!\circ\Lambda_{m+1}\! \Mcirc_{i=1}^{m} \!\left[\left(\!\mc{I}_\env\!\otimes\!\mc{G}_i^\dg\right)\!\circ\!\Lambda_i\!\circ\!\left(\mc{I}_\env\!\otimes\!\mc{G}_i\right)\right],
    \label{eq: non-Markov rb sequence}
\end{equation}
where $\mc{I}_\env$ is an identity map on $\env$, while $\Lambda$ is now a \gls{cptp} map on $\syst\env$ and $\mc{G}_i$ acts solely on $\syst$. The probabilities rendered by the \gls{asf} now correspond to Eq.~\eqref{eq: probabilities PT contraction} by replacing $\mc{U}_i\to\Lambda_i$ in $\Upsilon_m$, and replacing $\mc{A}_i\to\mc{G}_i^\dg\circ\mc{G}_{i-1}$ in $\Theta_m$, which is then averaged over each $\mc{G}_i$ gate.

As it will prove convenient, henceforth we will employ the superoperator~\cite{greenbaum2015} (a.k.a. Liouville~\cite{Wallman_2014} or natural~\cite{watrous2018theory}) representation of quantum channels, whereby quantum states get represented as vectors and quantum channels as matrices, both in spaces with respective dimensions squared. This is briefly detailed in Appendix~\ref{appendix: superoperator rep}.

\begin{notation} In particular, we distinguish the vectorized and the superoperator representation, respectively, by the double ket notation, $|\cdot\rrangle$, and by hats on maps, $\hat{\mc{X}}$.
\end{notation}

That is, the \gls{asf} for a non-Markovian \gls{rb} experiment is equivalently written as
\begin{align}
    \mc{F}_m &= \llangle{M}|\mbb{E}(\hat{\mc{S}}_m)|\rho\rrangle,\\
    \hat{\mc{S}}_m \!&= \hat{\tr}_\env\hat{\Lambda}_{m+1}\!\,\prod_{i=m}^1\!\left(\mbb1_\env\!\otimes\hat{\mc{G}}_i^\dg\right)\hat{\Lambda}_i\left(\mbb1_\env\!\otimes\hat{\mc{G}}_i\right), \label{eq: rb sequence nM superop}
\end{align}
where $\mbb1_\env$ is an identity on a $\dimE^2$-dimensional environment $\env$, $\hat{X}$ are $d^2\times{d}^2$ matrices (for either $d\sim\dimS$ or $d\sim\dimE\dimS$) and $|\cdot\rrangle$ is a $\dimE^2\dimS^2$ vector, with $\llangle\cdot|:=(|\cdot\rrangle)^\mathrm{T}$ being a co-vector. We will simply denote the spaces by $\env$ or $\syst$, with their dimensionality being implied by context.

In~\cite{figueroaromero2021randomised}, the exact functional form of the non-Markovian gate-independent \gls{asf} was computed for unitary 2-designs, and a set of methods was presented to estimate features of the noise. Despite this, such functional form remains somewhat obscure mathematically. Below we expand the class of gates considered to finite groups, and along we obtain an \gls{asf} which is much more transparent as to what the mechanism is behind the data of an \gls{rb} experiment subject to gate-independent non-Markovian noise.

\section{Non-Markovian Average Sequence Fidelity Beyond Unitary 2-Designs}\label{sec: asf}

We now relax the unitary 2-design restriction on the gates to be benchmarked to any finite-group admitting a multiplicity-free representation\footnote{ Relaxing this restriction can be done similarly as in~\cite{helsen2020general}; an in-depth analysis of this case might eventually be needed, given its experimental relevance, see e.g.~\cite{PRXQuantum.2.010351}.}. For all necessary background details on the representation of finite groups, we refer to Appendix~\ref{appendix: representations of finite groups}.

\begin{notation} Henceforth, we let $\mbb{G}$ be a finite subgroup of the $\dimS$-dimensional unitary group $\mbb{U}(\dimS)$, such that the superoperator representation of the $\mc{G}$ gates in Eq.~\eqref{eq: non-Markov rb sequence} and Eq.~\eqref{eq: rb sequence nM superop},
\begin{equation}
    \mc{\hat{G}}=\Moplus_{\pi\in{R}_\mathbb{G}}\phi_\pi(g)^{\otimes{n}_\pi},
    \label{eq: gates Maschke}
\end{equation}
is multiplicity-free, i.e., $n_\pi=1$ for all $\phi_\pi$, and where $\phi_\pi$ are the irreducible representations and $R_{\mbb{G}}$ is a set of labels for the corresponding irreducible subspaces.
\end{notation}

We begin with a proposition and a definition, both of which will make clearer the expression for the \gls{asf} of a \gls{rb} experiment under non-Markovian noise with gates sampled uniformly from the group $\mbb{G}$.

\begin{proposition}[Subspace $\phi$-twirl]\label{proposition: subsystem twirl}
    Let $\mc{T}_\phi$ denote the so-called twirl associated to the representation $\phi$ of $\mbb{G}$, i.e.
    \begin{equation}
    \mc{T}_\phi(\cdot) := \f{1}{|\mbb{G}|}\sum_{g\in\mbb{G}} \phi(g) (\cdot) \phi(g)^\dg,
    \end{equation}
    then, for a given \gls{cp} map $\Lambda$, we can write
    \begin{align}
    \left(\mc{I}_\env\otimes\mc{T}_\phi\right)\hat{\Lambda} = \sum_{\pi\in{R}_\mbb{G}}\left(\hat{\mc{Q}}_\pi\otimes\hat{\mc{P}}_\pi\right),
\end{align}
where $\hat{\mc{P}}_\pi$ is a projector operator onto the irreducible subspace defined by $\phi_\pi$, and
    \begin{gather}
        \hat{\mc{Q}}_\pi \!:=\!\!\! \sum_{e,e^\prime,\varepsilon,\varepsilon^\prime=1}^{\dimE} \!f_\pi^{ee^\prime\varepsilon\varepsilon^\prime}|ee^\prime\rangle\!\langle\varepsilon\varepsilon^\prime|,
        \label{eq: single quality map} \\
        \text{where}\qquad f_\pi^{ee^\prime\varepsilon\varepsilon^\prime} := \f{\tr\left(\langle{ee^\prime}|\hat{\Lambda}|\varepsilon\varepsilon^\prime\rangle\hat{\mc{P}}_\pi\right)}{\tr\left(\hat{\mc{P}}_\pi\right)}
    \end{gather}
with $\{|e\rangle\}_{e=1}^{\dimE}$, $\{|e^\prime\rangle\}_{e^\prime=1}^{\dimE}$ and $\{|\varepsilon\rangle\}_{\varepsilon=1}^{\dimE}$, $\{|\varepsilon^\prime\rangle\}_{\varepsilon^\prime=1}^{\dimE}$ all being arbitrary orthonormal bases for $\env$, and with identities on system $\syst$ being implicit (e.g., $\langle{e}|$ means $\langle{e}|\otimes\mbb1_\syst$).
\end{proposition}

The proof that Proposition~\ref{proposition: subsystem twirl} is true, can be seen in Appendix~\ref{appendix: nM finite group time-indep}, and it follows directly from Schur's lemma applied to the $\phi$-twirl.

In the absence of an environment, i.e., with $\Lambda=\Lambda^\markov$ being a noise map acting solely on $\syst$, this is $\mc{T}_\phi(\hat{\Lambda}^\markov)=\sum_\pi f_\pi \hat{\mc{P}}_\pi$, where
\begin{equation}
    f_\pi := \f{\tr(\hat{\Lambda}^\markov\hat{\mc{P}}_\pi)}{\tr(\hat{\mc{P}}_\pi)},
\end{equation}
was labeled a \emph{quality factor} in~\cite{Helsen2019} within the context of \gls{rb}. This motivates the following.

\begin{definition}[Quality maps]\label{def: quality maps}
Let us denote by $\vec{\epsilon}_n^{\,(\prime)}$ the sequence of $\env$-orthonormal basis labels $\epsilon_1$, $\epsilon_1^\prime$, $\epsilon_2$, $\epsilon_2^\prime$, \ldots, $\epsilon_{n-1}$, $\epsilon_{n-1}^\prime$. We call the map between $\env$ spaces,
    \begin{equation}
        \hat{\mc{Q}}_{n,\pi} := \sum_{e,e^\prime,\varepsilon,\varepsilon^\prime=1}^{\dimE} \mathbbm{f}_{n,\pi}^{ee^\prime\varepsilon\varepsilon^\prime} |ee^\prime\rangle\!\langle\varepsilon\varepsilon^\prime|,
        \label{eq: n quality map}
        \end{equation}
    the length-$n$ quality map associated to noise \gls{cp} maps $\Lambda_1,\Lambda_2,\ldots,\Lambda_n$ onto the $\pi$\textsuperscript{th} subspace defined by $\phi_\pi$, where
        \begin{align}
        \mathbbm{f}^{ee^\prime\varepsilon\varepsilon^\prime}_{n,\pi} \!\!:=\!\!\! \sum_{\{\epsilon_i,\epsilon_i^\prime=1\}}^{\dimE} \!\!\!\f{\tr\!\left(\langle ee^\prime\vec{\epsilon}_n^{\,(\prime)}|\!\otimes_i\!\hat{\Lambda}_i|\vec{\epsilon}_n^{\,(\prime)}\varepsilon\varepsilon^\prime\rangle\hat{\mc{P}}_\pi^{\otimes{n}}\right)}{\tr\left(\hat{\mc{P}}_\pi^{\otimes{n}}\right)},
        \end{align}
with sum over all $\epsilon_1,\epsilon_1^\prime,\ldots,\epsilon_{n-1},\epsilon_{n-1}^\prime$, is called an $n$-point quality factor, with $\hat{\mc{P}}_\pi$ being the projector operator onto the corresponding irreducible subspace.
\end{definition}

A length-1 quality map corresponds to that in Eq.~\eqref{eq: single quality map}, i.e. $\hat{\mc{Q}}_{n=1,\pi}=\hat{\mc{Q}}_\pi$, and such that $\mathbbm{f}_{n=1,\pi}^{ee^\prime\varepsilon\varepsilon^\prime}=f_{\pi}^{ee^\prime\varepsilon\varepsilon^\prime}$. For general $n$, the $n$-point quality factor $\mathbbm{f}^{ee^\prime\varepsilon\varepsilon^\prime}_{n,\pi}$ can be read as the components of an environment-dependent $n$-point function with correlations between maps $\Lambda_1,\ldots,\Lambda_n$, mediated through the environment via the $\epsilon_i^{(\prime)}$ components. This is because we may equivalently write the $n$-point quality factor as a product of single 1-point quality factors contracted through the environment, that is,
\begin{align}
    \mathbbm{f}_{n,\pi}^{ee^\prime\varepsilon\varepsilon^\prime} \!\!&=\!\! \sum_{\{\epsilon_i,\epsilon_i^\prime=1\}}^{\dimE}\!f_{(1),\pi}^{ee^\prime\epsilon_1\epsilon_1^\prime}f_{(2),\pi}^{\epsilon_1\epsilon_1^\prime\epsilon_2\epsilon_2^\prime}\cdots f_{(n),\pi}^{\epsilon_{n-1}\epsilon_{n-1}^\prime\varepsilon\varepsilon^\prime}\!\!,    
\end{align}
where $f_{(i),\pi}^{abcd} \!:= \tr(\langle{ab}|\hat{\Lambda}_i|cd\rangle\hat{\mc{P}}_\pi)/\tr(\hat{\mc{P}}_\pi)$, i.e., the subindex in parenthesis refers to the quality factor being associated to the $i$\textsuperscript{th} noise map $\Lambda_i$, and sum is over all $\epsilon_i$ and $\epsilon_i^\prime$ indices, with $e,e^\prime,\varepsilon,\varepsilon^\prime$ indices being free (summed over in Eq.~\eqref{eq: n quality map}).

In the absence of an environment, with $\Lambda^\markov_1,\ldots,\Lambda^\markov_m$ acting solely on $\syst$, the quality map turns into a scalar $\mc{Q}_{m,\pi}=\prod_i f_{(i),\pi}$, product of $m$ quality factors $f_{(i),\pi}=\tr(\hat{\Lambda}^\markov_i\hat{\mc{P}}_\pi)/\tr(\hat{\mc{P}}_\pi)$, which are now all independent of each other. In the context of \gls{rb}, this renders the \gls{asf} as $\mc{F}_m^\markov=\sum_\pi\prod_if_{(i),\pi}\llangle{M}|\hat{\Lambda}^\markov_{m+1}\hat{\mc{P}}_\pi|\rho_{_\syst}\rrangle$ for time-dependent noise, or as a linear combination of exponentials $\mc{F}_m^\markov=\sum_\pi f_\pi^m\llangle{M}|\hat{\Lambda}^\markov\hat{\mc{P}}_\pi|\rho_{_\syst}\rrangle$, as in~\cite{Helsen2019, hashagen_finite}, for time-independent noise. For unitary 2-designs, as detailed in Appendix~\ref{appendix: Markov 2 design}, there are two invariant subspaces with ${f}_{\pi=1}=1$ and $f_{\pi=2}=(\tr[\hat{\Lambda}^\markov]-1)/(\dimS^2-1)$, with the remaining $\llangle{M}|\hat{\Lambda}^\markov\hat{\mc{P}}_\pi|\rho_{_\syst}\rrangle$ corresponding to the \gls{spam} error constants $A$ and $B$. The trace-preserving property of the noise gives rise to the constant unity factor $f_{\pi=1}$ for the trivial subspace; otherwise, for trace non-increasing noise, this quality factor corresponds to a trace-loss quantifier.

Still within the Markovian case, as explained in~\cite{Helsen2019}, for non-Clifford gate sets, the quality factors do not always have the same straightforward interpretation as a noise strength or trace-loss, precisely because their contributions to the \gls{asf} end up in different irreducible subspaces, although these still generally provide information about the quality of the noisy gates. Turning to the non-Markovian case, we first present the following.

\begin{theorem}[Average Sequence Fidelity]\label{thm: asf} Given a non-Markovian \gls{rb} sequence $\hat{\mc{S}}_m$ of length $m$ as in Eq.~\eqref{eq: rb sequence nM superop}, with the gates $\mc{G}_i$ satisfying Eq.~\eqref{eq: gates Maschke} with irreducible representations $\phi_\pi$, and $\Lambda_i$ being \gls{cp} maps on $\syst\env$, the corresponding \gls{asf} with an initial state $\rho$ and measurement \gls{povm} element $M$ is given by
\begin{equation}
    \mc{F}_m = \sum_{\pi\in{R}_{\mbb{G}}}\! \llangle{M}|\hat{\tr}_\env\hat{\Lambda}_{m+1}\left(\hat{\mc{Q}}_{m,\pi}\otimes\hat{\mc{P}}_\pi\right)|\rho\rrangle,\label{eq: main asf}
\end{equation}
where $R_\mbb{G}$ is set of labels for the spaces associated to each irreducible representation, the $\hat{\mc{P}}_\pi$ are projector operators onto these, and $\mc{Q}_{m,\pi}$ is the length-$m$ quality map associated to the maps $\Lambda_1,\Lambda_2,\ldots,\Lambda_m$ on the $\pi$\textsuperscript{th} irreducible subspace, as per Definition~\ref{def: quality maps}.
\end{theorem}
The proof follows directly from Proposition~\ref{proposition: subsystem twirl} and Definition~\ref{def: quality maps} but is presented coherently as well in Appendix~\ref{appendix: nM finite group time-indep}.

Despite dealing with a more general and abstract scenario than that of the unitary 2-design case, the \gls{asf} in Theorem~\ref{thm: asf} is rather conceptually simple in that information about intermediate noise, including all its correlations, is carried through the length-$m$ quality maps $\mc{Q}_{m,\pi}$ through the environment across the $\pi$ irreducible subspaces. Similar to the Markov case, these subspaces might be thought of as quality sectors for the noise, being the trivial subspace the corresponding to the ideal noiseless case. As shown in detail in Appendix~\ref{appendix: asf nM Clifford}, the unitary 2-design case reduces to the expression obtained in~\cite{figueroaromero2021randomised}, of the form $\mc{F}_m=\llangle{M}|\hat{\tr}_\env\hat{\Lambda}_{m+1}|(\mscr{A}_m+\mscr{B}_m)\rho\rrangle$, where the maps $\mscr{A}_m$ and $\mscr{B}_m$ contain corresponding quality maps related to the depolarizing effect of the average over system $\syst$.

Now, given that in realistic scenarios temporal correlations are effectively finite~\cite{PhysRevLett.122.140401, PhysRevA.99.042108}, a relevant case is that when the \gls{asf} contains only a smaller length with non-Markovian behavior than that of the full sequence length. We then have the following.

\begin{corollary}[Finite non-Markovianity]\label{Thm: non-Markov perturbative asf} Let
\begin{equation}
    \Lambda = q \Gamma + (1-q) \Phi^\markov,\,\text{with}\,\,\,0\leq{q}\leq1,
    \label{eq: model finite noise}
\end{equation}
where $\Gamma$ is a \gls{cp} map between $\syst\env$ spaces and $\Phi^\markov$ is a \gls{cp} map solely between $\syst$ spaces. Then, for a \gls{rb} sequence of length $m$, with time-independent noise, i.e., $\Lambda_i=\Lambda_j$ for all $i\neq{j}$, we have
\begin{align}
    \mc{F}_m &= (1-q)^m\mc{F}_m^\markov\nonumber\\ &\quad + \sum_{\ell=1}^{m} \binom{m}{m-\ell}q^\ell(1-q)^{m-\ell}\mc{F}_m^{(m-\ell)}, 
    \label{eq: asf perturb nM}
\end{align}
where $\binom{a}{b}:=\f{a!}{b!(a-b)!}$ is a binomial coefficient, and where
\begin{align}
    \mc{F}_m^{(m-\ell)} \!&:=\!\! \sum_{\pi\in{R}_{\mathbb{G}}}\!\!f_\pi^{m-\ell}\!\llangle{M}|\hat{\tr}_\env\hat{\Lambda}\!\left(\!\hat{\mc{Q}}_{\ell,\pi}\!\otimes\!\hat{\mc{P}}_\pi\!\right)\!|\rho\rrangle,
    \label{eq: asf finite memory}
\end{align}
with $f_\pi$ being quality factors associated to $\Phi^\markov$ and $\mc{Q}_{n,\pi}$ the quality map associated to $n$ copies of $\Gamma$. We denote by $\mc{F}_m^\markov=\mc{F}_m^{(m)}$ the fully Markovian \gls{asf}.
\end{corollary}

The proof is shown in detail in Appendix~\ref{appendix: finite memory asf} for the more general time-dependent case, where similarly the noise at the $i$\textsuperscript{th} step is modeled as $\Lambda_i=q_i\Gamma_i+(1-q_i)\Phi_i^\markov$ with $0\leq{q}_i\leq1$ and $\Lambda_i\neq\Lambda_j$ for all $i\neq{j}$.

In particular, Eq.~\eqref{eq: asf perturb nM} becomes a relevant perturbative non-Markovian expansion whenever $0\lesssim{q}\ll{1/2}$. This may serve to analyze finite non-Markovian deviations on smaller sequence length intervals whenever a model of the form in Eq.~\eqref{eq: model finite noise} is available. In any case, Eq.~\eqref{eq: asf finite memory} describes a finite-memory \gls{asf} with $\ell$ steps displaying non-Markovian deviations; this serves as a generalization of the finite-memory case for 2-designs introduced in~\cite{figueroaromero2021randomised}, and similarly may be used to estimate the Markov order, as well as several related features, for finite-memory effects in quantum devices.

In general, the central quantity capturing average noise in non-Markovian \gls{rb} remains the quality maps $\mc{Q}_{m,\pi}$. In the following section we propose a way to operationally extract information about these through a slight modification to the standard \gls{rb} protocol.

\section{Average Process Fidelity}\label{sec: avg gate fidelity}
The main advantage of \gls{rb} is that decays of the \gls{asf} can be straightforwardly estimated experimentally; in particular in the Markovian case, obtaining relevant figures of merit, such as the average gate fidelity of the physical gates relative to the ideal ones, is reduced to a fitting problem. In the case of finite groups, the fitting has to be done to a linear combination of exponentials: this can be achieved via a modified \gls{rb} protocol called Character Randomized Benchmarking~\cite{Helsen2019}, allowing to operationally estimate the individual decays over each irreducible subspace (including the non multiplicity-free case~\cite{PRXQuantum.2.010351, helsen2020general}). While in the non-Markovian case it is possible to execute an analogous Character \gls{rb} protocol, as pointed out in Appendix~\ref{appendix: character rb}, the fact that the individual irreducible parts of the \gls{asf} are a non-trivial function of the environment, remains.

In the case of Markovian, local noise in system $\syst$, the so-called average gate fidelity of physical gates, which we can model as $\Lambda^\markov\circ\mc{C}_i$, with respect to the ideal gates $\mc{C}_i$, is equivalent to that of $\Lambda^\markov$ with respect to the identity $\mc{I}_\syst$, which is defined as
\begin{equation}
    \mathfrak{F}_{\Lambda^\markov} := \int d\psi\, \langle\psi|\Lambda^\markov(|\psi\rangle\!\langle\psi|)|\psi\rangle,
    \label{eq: avg gate fidelity}
\end{equation}
with pure states $|\psi\rangle\in\mscr{H}_\syst$. As in~\cite{Helsen2019}, it can be shown that this gate fidelity is related to the data outputs of the \gls{rb} protocol as
\begin{equation}
    \mathfrak{F}_{\Lambda^\markov} = \f{\dimS+\sum_{\pi\in{R}_{\mbb{G}}}f_\pi\tr(\hat{\mc{P}}_\pi)}{\dimS(\dimS+1)},
\end{equation}
where $f_\pi$ here are the quality factors of $\Lambda^\markov$, which can be estimated individually through Character \gls{rb}.

For the non-Markovian case, however, the quality map is dependent on the environment $\env$, and furthermore, \gls{spam} errors might correlate $\syst$ with $\env$, thus affecting the final fidelity outputs. This stresses the need of a figure of merit benchmarking full quantum noise processes rather than individual noise rates whenever errors are temporally-correlated. In a sense, too, it is rather error correlations within \gls{spam} which matter, rather than the local errors themselves.

Frameworks such as~\cite{helsen2021estimating} and~\cite{flammia2021averaged}, generalizing the \gls{rb} theoretical framework and technique have been proposed which could provide new ideas to tackle this problem. Here we point out the following. Let us define
\begin{align}
    \hat{\fid}_{m,\pi}^\syst|\!\cdot\rrangle &:=  \hat{\tr}_\env\hat{\Lambda}_{m+1}\!\left(\hat{\mc{Q}}_{m,\pi}\!\otimes\hat{\mc{P}}_\pi\right)\!\hat{\Lambda}_0|\varepsilon\otimes\!\cdot\rrangle,
    \label{eq: def fid map}
\end{align}
where here $\Lambda_0$, acting on the full $\syst\env$, encodes state preparation errors and correlations, for some fiducial pure state $\varepsilon$ of $\env$, and so that Eq.~\eqref{eq: main asf} turns into $\mc{F}_m=\sum_\pi\llangle{M}|\hat{\fid}_{m,\pi}^\syst|\rho_{_\syst}\rrangle$ for some prepared initial state $\rho_{_\syst}$ of $\syst$. We can regard the output data of the \gls{asf} as a distribution in initial states $\rho_{_\syst}$ and measurements $M$, which are the parameters we can still fix; further averaging over initial states and \gls{povm} elements is equivalent to obtaining the average gate fidelity of $\sum_\pi\fid_{m,\pi}^\syst$, and individual instances for each subspace, $\fid_{m,\pi}^\syst$, can be estimated via Character \gls{rb}.

As per the definition in Eq.~\eqref{eq: avg gate fidelity}, for simplicity, we may consider first randomizing both the initial state and the measurement element with a fixed unitary map $\mc{U}$, drawn uniformly at random from a unitary 2-design, so that averaging over it gives the average gate fidelity of the map $\fid_{m,\pi}^\syst$ with respect to the identity. That is, we take
\begin{subequations}
\begin{align}
   & \rho_{_\syst} = \mc{N}\circ\mc{U}(|\psi\rangle\!\langle\psi|) \\
   & M_r = \mc{M}\circ\mc{U}(|r\rangle\!\langle{r}|),
\end{align}
\end{subequations}
where $\mc{N}$ and $\mc{M}$ are noise, \gls{cp} maps, and $\mc{U}$ is a unitary map, all acting solely between $\syst$ spaces, and with $|\psi\rangle$, $|r\rangle$ an arbitrary vector and orthonormal basis vector of $\syst$, respectively. The initial state $\rho_{_\syst}$ is now a random initial noisy state with pure target state $|\psi\rangle$, and $M_r$ is the $r$\textsuperscript{th} element of the random \gls{povm}~\cite{heinosaari2019random} $\{M_r\}$. The local \gls{spam} noise, $\mc{N}$ and $\mc{M}$, which we take as independent of $\mc{U}$, are due to randomization and can now simply be absorbed in $\Lambda_0$ and $\Lambda_{m+1}$ in Eq.~\eqref{eq: def fid map}. This could equivalently be done by drawing the gates $\mc{G}_1=\mc{G}_m=\mc{U}$ on each run of the \gls{rb} protocol from a unitary 2-design.

As detailed in Appendix~\ref{appendix: process fidelity}, averaging over such initial states and measurements, and taking the noise to be trace-preserving, we get
\begin{align}
    \mbb{E}_{\rho_{_\syst},M_r}\left[\mc{F}_m\right] = A \mathfrak{F}_{\fid}^{(m)} + B,
    \label{eq: avg states measurements gate fidelity}
\end{align}
where
\begin{equation}
    \mathfrak{F}_{\fid}^{(m)} = \f{\dimS + \sum_\pi\tr\left(\hat{\fid}_{m,\pi}^\syst\right)}{\dimS(\dimS+1)},
    \label{eq: avg process fidelity}
    \end{equation}
and
\begin{equation}
    A =1-\dimS B,\qquad B = \f{1-|\langle{r}|\psi\rangle|^2}{\dimS-1},
\end{equation}
which directly gives an average gate fidelity for the full $m$-step noisy process, taking into account all correlations within, including those induced by the \gls{spam} noise. Ideally, the target initial state and measurement should satisfy $\langle{r}|\psi\rangle=1$, so that $\mbb{E}_{\rho_{_\syst},M_r}\left[\mc{F}_m\right] = \mathfrak{F}_{\fid}^{(m)}$; otherwise the only problematic case is if these are mutually unbiased, $|\langle{r}|\psi\rangle|^2=1/\dimS$, where no information about the average gate fidelity is obtained.

In particular, Eq.~\eqref{eq: avg process fidelity} is such that
\begin{equation}
    \tr\left(\hat{\fid}_{m,\pi}^\syst\right) \leq \|\hat{\mscr{S}}\| \tr\left(\hat{\mc{Q}}_{m,\pi}\right)\tr\left(\hat{\mc{P}}_\pi\right),
    \label{eq: bound spam-quality}
\end{equation}
where $\hat{\mscr{S}}:=\hat{\Lambda}^\prime_0\hat{\mathfrak{S}}_{\varepsilon}\hat{\tr}_\env\hat{\Lambda}^\prime_{m+1}$ is a \gls{spam}-only operator, here with $\hat{\mathfrak{S}}_{\varepsilon}:=(|\varepsilon\rangle\otimes\mbb1)\otimes(|\varepsilon\rangle\otimes\mbb1)^*$ and primed operators, $\Lambda_0^\prime$ and $\Lambda^\prime_{m+1}$, indicating the absorbed $\mc{N}$ and $\mc{M}$ local \gls{spam} noise, respectively. While Eq.~\eqref{eq: avg states measurements gate fidelity} already gives an operationally meaningful average gate fidelity for a full \gls{rb} process --more rightly called an \emph{average process fidelity--,} one may estimate separately the noise influence from \gls{spam}-induced correlations from that within the quality map of the process via Character \gls{rb} and Eq.~\eqref{eq: bound spam-quality}, albeit requiring a prior estimate for the weight of the \gls{spam} noise.

Whenever the noise is effectively Markovian, clearly, averaging over initial states and measurements simply amounts to averaging the \gls{spam} terms; while redundant when only the gate set is of interest, this could also serve to estimate systematic average \gls{spam} error rates. The average process fidelity in Eq.~\eqref{eq: avg process fidelity} for the Markov case simply turns the trace into a product of quality parameters, including that of the \gls{spam} noise; for the unitary 2-design case these can be related back to individual average gate fidelities, and the \gls{asf} is simply $\mbb{E}_{\rho_{_\syst},M_r}[\mc{F}_m]=\alpha p^mq+\beta$ for $\alpha$, $\beta$ constants just depending on the inner product of targets $|\langle{r}|\psi\rangle|^2$, and $q$ the average noise-strength of the \gls{spam} noise, as $q=(\tr[\hat{\Lambda}_0\hat{\Lambda}_{m+1}]-1)/(\dimS^2-1)$. This is worked out in detail in Appendix~\ref{appendix: avg process fidelity Markov}.

A crucial difference now, is that not only the decay profile of the \gls{asf}, but also the average fidelity for the whole process is dependent non-trivially on the sequence length $m$ through the environment via the quality factors $\mc{Q}_{m,\pi}$. Then again, this is a consequence of the presence of multi-time correlations. While this implies that it is not possible to fit a simple function of sequence length to the \gls{asf}, it can serve as a diagnosis for non-Markovianity and a means to quantify it, as discussed next.

\section{Signature of non-Markovian noise in Randomized Benchmarking}\label{sec: nM}
A first signature of non-Markovianity in a \gls{rb} experiment is the display of deviations from an exponential decay in the \gls{asf}~\cite{figueroaromero2021randomised}. However, deviations might not be evident or significant, either statistically after short sequence lengths, or fundamentally due to the noise itself, making the experiment blind to non-Markovianity, or the noise might be time-dependent but Markovian, not displaying multi-time correlations but rather some arbitrary temporal dependence changing the exponential rates of decay of the \gls{asf} data. Similarly, gate-dependence generally generates deviations from an exponential~\cite{PhysRevLett.119.130502,Wallman_2014}. Anyhow, using the framework established before, we may look for unique signatures of non-Markovianity.

In~\cite{Wallman_2014} it is pointed out that, in the context of \gls{rb}, \emph{if some of the} (quality) \emph{parameters are observed to be greater than 1, the experimental noise must be non-Markovian}. Wallman and Flammia notice that the decay of the \gls{asf} for time-dependent Markovian noise is given by products of time-dependent quality parameters, so that these can only change the rate of decay, as all of them are upper-bounded by unity. Having an increase within the \gls{asf} in increasing sequence length thus points to non-Markovianity. As can be seen with the analysis presented above, the same will hold for \gls{rb} with the finite group $\mbb{G}$, with the decay specified by $f_{(1),\pi}\cdots{f}_{(m),\pi}$, and with each quality factor satisfying $f_{(i),\pi}=\tr(\hat{\Lambda}\hat{\mc{P}}_
\pi)/\tr(\hat{\mc{P}}_\pi)\leq1$ for all $\pi$, due to $\Lambda$ being trace non-increasing, or in particular trace-preserving.

That is, in general having $\mc{F}_n>\mc{F}_m$ for any $n>m$ must imply a non-Markovian effect, as this cannot be explained within the Markovian framework~\footnote{ The case of gate-dependence, discussed in the following section, can be regarded as a form of non-Makovianity.}. Indeed, in the following lemma we give sufficient and necessary conditions, which can only be satisfied in a non-Markovian framework, for the condition $\mc{F}_n>\mc{F}_m$ with $n>m$ to occur.

\begin{lemma}[Conditions for non-monotonic \gls{asf}] Let $n,m$ be positive integers such that $n>m$, and let $\mc{F}_n$ and $\mc{F}_m$ be two \gls{asf} corresponding to the same underlying noise process described by \gls{cptp} maps $\Lambda_1,\ldots,\Lambda_m$ and $\Lambda_1,\ldots,\Lambda_n$, respectively, acting on $\syst\env$. A sufficient condition for $\mc{F}_n > \mc{F}_m$, is
\begin{align}
     \hat{\Lambda}_{n+1} \left(\hat{\mc{Q}}_{n,\pi}\otimes\mbb1\right) \succ \hat{\Lambda}_{m+1}\left(\hat{\mc{Q}}_{m,\pi}\otimes\mbb1\right),
    \label{eq: quality nM positive}
\end{align}
for all $\pi$, where $\mc{Q}_{m,\pi}$ and $\mc{Q}_{n,\pi}$ are corresponding quality maps to $\mc{F}_m$ and $\mc{F}_n$, respectively; i.e., that the difference of matrix quality maps, $\hat{\Lambda}_{n+1} (\hat{\mc{Q}}_{n,\pi}\otimes\mbb1) - \hat{\Lambda}_{m+1}(\hat{\mc{Q}}_{m,\pi}\otimes\mbb1)$, is positive definite. Furthermore, the condition
\begin{align}
    \tr\left(\hat{\Lambda}_{n+1}\right)\prod_{i=m+1}^{n}\|\hat{\Lambda}_i\| &> \dimS\tr\left(\hat{\Lambda}_{m+1}\right),
    \label{eq: opnorm Lambdas}
\end{align}
is necessary, where $\|\cdot\|$ denotes operator norm, here corresponding to maximum singular value.
\end{lemma}

The proof is shown in Appendix~\ref{appendix: non-Markov}, mainly relying on the inequalities $\tr(XY)\leq\|X\|\tr(Y)$ for positive $X$, $Y$, and $\|\tr_A(X_{AB})\|\leq{d}_{\mathsf{A}}\|X_{AB}\|$ proven in~\cite{Rastegin}. While Eq.~\eqref{eq: quality nM positive} is an expected consequence of all information about the average noise being carried within the quality maps, Eq.~\eqref{eq: opnorm Lambdas} places a relevant necessary constraint on the noise represented by the $\Lambda_i$ maps. Clearly, conditions in Eq.~\eqref{eq: quality nM positive} and Eq.~\eqref{eq: opnorm Lambdas} cannot be satisfied within Markovianity, i.e., if the quality maps are quality factors and if the noise maps act solely on subsystem $\syst$.

In particular, consider time-independent noise, $\Lambda_i=\Lambda_j=\Lambda$ for all $i\neq{j}$, and $n=m+1$, then Eq.~\eqref{eq: opnorm Lambdas}, now $\|\hat{\Lambda}\|>\dimS$, says that the maximum increase of purity by the map $\Lambda$ on the full $\syst\env$ state at the corresponding step must be over $\dimS^2$, which already rules out, for example, coherent (unitary) noise. For an increase in the \gls{asf} after a sequence length $n-m>0$ with time-independent noise, we get $\|\hat{\Lambda}\|>\dimS^{1/(n-m)}$. Moreover, this is relevant as well using the inequality $\|\hat{\mc{X}}\|\leq\sqrt{d}$ for any \gls{cptp} map acting between $d$-dimensional spaces (Theorem II.I in~\cite{OpNormCPTP_2006}), as it implies that the environment needs $\dimE>\dimS^{2/(n-m)-1}$ to witness such $\mc{F}_n>\mc{F}_m$ with $n>m$. In general, as the trace terms in Eq.~\eqref{eq: opnorm Lambdas} are normally in practice different from zero, and almost always close to $\dim_{\env\syst}^2$, this condition can be interpreted analogously, as requiring an increase of purity by the noise of at least $\dimS^2\left(\tr\hat{\Lambda}_{m+1}/\tr\hat{\Lambda}_{n+1}\right)^2$, with the ratio of traces being close to unity.

While Markovian time-dependence can be obtained via a superfluous environment getting discarded between steps, non-Markovianity implies a time-dependence such that an environment correlates timesteps with each other. In particular, with Markovian time-dependence, one can estimate average gate fidelities over arbitrary sequence length intervals, as also shown in~\cite{Wallman_2014}, because temporal modularity is such that products of quality factors satisfy $(f_{1}\cdots{f}_{n})/(f_{1}\cdots{f}_{m})=f_{m+1}\cdots{f}_{n}$ for any $n>m$ (and for all $\pi$, which we omitted here). An analogous property is not satisfied by quality maps, precisely due to any given step depending on all previous ones.


Understanding such unique non-Markovian deviations more deeply could be relevant to further be able to use this to one's own advantage, either for control, mitigation or otherwise, of the noise in question. Up to this point, however, we have assumed gate-independence for the noise: a remaining question is thus, what the impact of gate-dependence, and/or context errors, is when considered together with non-Markovianity?
\section{Gate-dependence in non-Markovian Randomized Benchmarking}\label{sec: gate-dependence}
In~\cite{PhysRevLett.119.130502, Wallman_2018}, it was shown that taking into account gate-dependence within the noise, whenever we assume Markovianity, the \gls{asf} for a \gls{rb} experiment with unitary 2-designs will behave as an exponential plus a perturbative term due to gate-dependence, which itself decays exponentially in sequence-length. This result is further extended in~\cite{Helsen2019} to \gls{rb} with the group $\mbb{G}$ as we have considered here, with the same conclusion. Further analyses in more generality can also be seen in Ref.~\cite{merkel2018randomized, helsen2020general}. This result, however, does not extend to the case of non-Markovian noise.

Here, to make this point\footnote{ A very detailed analysis, including an in-depth discussion of~\cite{PhysRevLett.119.130502, Wallman_2018} (whose interpretation is seen to be equivalent), can be seen in~\cite{helsen2020general}, including a further discussion of precisely \emph{what \gls{rb} actually measures} (also discussed and answered in~\cite{merkel2018randomized}), which we do not address here.}, we follow the argument in~\cite{Wallman_2018}, which starts by noticing that instead of using the $\Lambda$ maps, we could have instead chosen to model the noisy gates as $\mc{J}:=\mc{L}\circ\mc{G}\circ\mc{R}$ for \gls{cp} maps $\mc{L}$ and $\mc{R}$; this would render an equivalent \gls{asf} to the standard Markovian one with unitary 2-designs. Having gate-dependent noise means either $\mc{L}$, $\mc{R}$, or both, depend on $\mc{G}$, and we can write, e.g., $\mc{J}^{(g)}:=\mc{L}_g\circ\mc{G}\circ\mc{R}$. More generally, then, we may define the map $\Delta_g := \mc{J}^{(g)} + \mc{J}$ capturing all gate-dependence in the noisy gates. Denoting $X_{j:i} = X_j\cdots{X}_i$, we can expand the corresponding noisy sequence as
\begin{align}
    &\hat{\mc{S}}_m^{(g)} = \hat{\mc{J}}^{(g)}_{m+1:1} \nonumber\\
    &= \sum_{\{\ell_i=0\}}^1 \hat{\mc{J}}_{m+1}^{\ell_{m+1}}\hat{\Delta}_{m+1}^{1-\ell_{m+1}}\cdots\hat{\mc{J}}_{1}^{\ell_1}\hat{\Delta}_{1}^{1-\ell_1} \nonumber\\
    &= \hat{\mc{J}}_{m+1:1} + \hat{\Delta}_{m+1:1} + \kern-2.5em \sum_{\substack{\{\ell_i=0\}\\[0.25em]\setminus\{\text{all}\,\ell_i=0\,\text{or}\,\text{all}\,\ell_i=1\}}}^1 \kern-2.5em \left(\prod_{i=m+1}^{1}\!\!\!\hat{\mc{J}}_{i}^{\ell_i}\hat{\Delta}_{i}^{1-\ell_i}\!\right),
    \label{eq: sequence gate dep}
\end{align}
where here $\Delta_i=\Delta_{g_i}$, and where where the second line follows from the (multi) binomial theorem. Of particular relevance is the rightmost term in the last line in Eq.~\eqref{eq: sequence gate dep}, which mixes $\mc{J}$ and $\Delta$ terms: the result in~\cite{Wallman_2018} (and generalization in~\cite{Helsen2019}) imply that for Markovian noise, such mixed terms do not contribute to the \gls{asf}, and that the term $\Delta_{m+1:1}$ gives a contribution that vanishes exponentially in sequence length.

Generalizing this result with the group $\mbb{G}$, as in~\cite{Helsen2019}, but when the noise is non-Markovian, would require for there to be $\mc{L}$ and $\mc{R}$ satisfying the properties
\begin{subequations}\label{eq: gate-dep}
\begin{align}
    \mbb{E}\left[\hat{\mc{J}}^{(g)}\hat{\mc{L}}\,\hat{\mc{G}}^\dg\right] &= \hat{\mc{L}}\,\hat{\mc{D}}_g,\label{eq: gate dep left}\\
    \mbb{E}\left[\hat{\mc{G}}^\dg\hat{\mc{R}}\,\hat{\mc{J}}^{(g)}\right] &= \hat{\mc{D}}_g \hat{\mc{R}},\label{eq: gate dep right}\\
    \mbb{E}\left[\hat{\mc{G}}\,\hat{\mc{R}}\hat{\mc{L}}\,\hat{\mc{G}}^\dg\right] &= \hat{\mc{D}}_g,\label{eq: gate dep middle}
\end{align}
\end{subequations}
where $\mbb{E}$ here again is uniform averaging over the group $\mathbb{G}$, and with
\begin{equation}
    \hat{\mc{D}}_g := \sum_{\pi\in{R}_\mathbb{G}} \hat{\mscr{Q}}_\pi\otimes \hat{\mc{P}}_\pi,
\end{equation}
for some (length-1) quality map $\mscr{Q}_\pi$, and where $\hat{\mc{P}}_\pi$ is a projector onto the irreducible subspace defined by the representation $\phi_\pi$ of $\mc{G}$ as in Eq.~\eqref{eq: gates Maschke}.

Following~\cite{Helsen2019}, we start by plugging the definition of $\mc{D}_g$ in Eq.~\eqref{eq: gate dep left} and Eq.~\eqref{eq: gate dep right}, together with the multiplicity-free decomposition in Eq.~\eqref{eq: gates Maschke}, so that
\begin{subequations}
\begin{align}
    \sum_\pi\mbb{E}\!\left[\hat{\mc{J}}^{(g)}\hat{\mc{L}}\hat{\mc{P}}_\pi\phi_\pi(g)^\dg\right] \!&=\! \sum_\pi\hat{\mc{L}}\!\left(\!\hat{\mscr{Q}}_\pi\!\otimes\! \hat{\mc{P}}_\pi\!\right),\\
    \sum_\pi\mbb{E}\!\left[\phi_\pi(g)^\dg\hat{\mc{P}}_\pi\hat{\mc{R}}\hat{\mc{J}}^{(g)}\right] \!&=\! \sum_\pi \!\left(\!\hat{\mscr{Q}}_\pi\!\otimes \hat{\mc{P}}_\pi\!\right)\!\hat{\mc{R}}.
\end{align}
\end{subequations}

Now both $\hat{\mc{P}}_\pi$ and $\phi(g)$ act solely on $\syst$, so we now take without loss of generality,
\begin{subequations}
\begin{align}
    \hat{\mc{L}}:=\sum_\pi\hat{\mc{L}}_\pi,\quad \hat{\mc{L}}_\pi\left(\mbb1_\env\otimes\hat{\mc{P}}_\lambda\right) = \delta_{\pi\lambda}\hat{\mc{L}}_\pi, \label{eq: L decomp nM}\\
    \hat{\mc{R}}:=\sum_\pi\hat{\mc{R}}_\pi,\quad \left(\mbb1_\env\otimes\hat{\mc{P}}_\lambda\right)\,\hat{\mc{R}}_\pi = \delta_{\pi\lambda}\hat{\mc{R}}_\pi,\label{eq: R decomn nM}
\end{align}
\end{subequations}
both for all $\lambda$, and hence, writing corresponding identities explicitly, we have the equations
\begin{subequations}
\begin{align}
    \mbb{E}\left[\hat{\mc{J}}^{(g)}\hat{\mc{L}}_\pi\left(\mbb1_\env\otimes\phi_\pi(g)^\dg\right)\right] &= \hat{\mc{L}}_\pi\left(\hat{\mscr{Q}}_\pi\otimes\mbb1_\syst\right), \\
    \mbb{E}\left[\left(\mbb1_\env\otimes\phi_\pi(g)^\dg\right)\hat{\mc{R}}_\pi\hat{\mc{J}}^{(g)}\right] &= \left(\hat{\mscr{Q}}_\pi\otimes\mbb1_\syst\right)\hat{\mc{R}}_\pi,
\end{align}
\end{subequations}
and we can vectorize both sides\footnote{ By means of ${\mathrm{vec}(AXB)=\left(A\otimes{B}^\mathrm{T}\right)\mathrm{vec}(X)}$; this follows from the definition we employ here, $\mathrm{vec}\left(|i\rangle\!\langle{j}|\right):=|ij\rangle$.}, and reorder spaces to get
\begin{subequations}\label{eq: quality eval general env}
\begin{align}
    &\llangle\hat{\mc{L}}_\pi|\left\{\mbb1_\env\otimes\mbb{E}[(\phi_\pi(g)^*\otimes\hat{\mc{J}}^{(g)})^{\mathrm{T}}]\right\} \nonumber\\
    &\hspace{7em}= \llangle\hat{\mc{L}}_\pi|(\hat{\mscr{Q}}_\pi\otimes\mbb1_{\syst\env\syst}), \\
    &\left\{\mbb1_\env\otimes\mbb{E}[(\phi_\pi(g)^*\otimes\hat{\mc{J}}^{(g)})^{\mathrm{T}}]\right\} |\hat{\mc{R}}_\pi\rrangle \nonumber\\
    &\hspace{7em}= (\hat{\mscr{Q}}_\pi\otimes\mbb1_{\syst\env\syst})|\hat{\mc{R}}_\pi\rrangle.
\end{align}
\end{subequations}

At this point, for Markovian noise, the quality map $\mscr{Q}_\pi$ is just a quality factor $f_\pi$, thus giving right and left eigenvalue equations for the operator $\mbb{E}[(\phi_\pi(g)^*\otimes\hat{\mc{J}}^{(g)})^{\mathrm{T}}]$, and unfolding the proof for an existence of $\mc{L}$, $\mc{R}$ operators satisfying all Eq.~\eqref{eq: gate-dep}. This is not possible in general for the non-Markovian case, simply because of the presence of $\env$. Furthermore, Eq.~\eqref{eq: quality eval general env} make manifest that average gate-dependent errors will get carried within $\env$ through quality maps.

We may notice, however, that if the gate-dependence is small in the sense of the physical implementation being given by
\begin{align}
    \mc{J}^{(\epsilon,g)} := \Lambda\circ\mc{G} + \epsilon \Delta_g,
\end{align}
for an $\epsilon$ scaled such that $\|\hat{\Delta}_g\|\leq1$, then the \gls{asf} and its variance remain close to the unperturbed ones. This can be done as in~\cite{Wallman_2014}, as shown in Appendix~\ref{appendix: gate-dep stability}, so that $|\mc{F}_m-\mc{F}_m^{(\epsilon,g)}|\leq\delta_\mc{F}$, where $\mc{F}_m^{(\epsilon,g)}$ is the \gls{asf} corresponding to a sequence of perturbed gates $\mc{J}^{(\epsilon,g)}_{m+1:1}$, whenever
\begin{align}
    \epsilon \lesssim \f{\delta_{\mc{F}}}{(m+1)\dimE\dimS},
\end{align}
where in particular, with non-Markovian noise, the scaling with increasing $\dimE$ also requires a smaller $\epsilon$, in a sense also pointing out the relevance of the environment for general, not necessarily small, gate-dependence. For the variance, similarly, if for some $\delta_{\mc{V}}>0$,
\begin{align}
    \epsilon \lesssim \f{\delta_{\mc{V}}}{4(m+1)\dimE\dimS},
\end{align}
then this implies $|\mc{V}_m - \mc{V}_m^{(\epsilon,g)}|\leq\delta_{\mc{V}}$, where $\mc{V}_m$ and $\mc{V}_m^{(g)}$ are variances of the unperturbed and perturbed sequence fidelities, respectively. This bound would thus require a perturbation a quarter times smaller for a change in variance of the same magnitude as a change in average of the sequence fidelity. Both bounds are almost the same as in~\cite{Wallman_2014}, with the extra $\dimE$ factor being a consequence of the noise acting on the whole $\syst\env$; the fact that \gls{rb} is done only on subsystem $\syst$ is of no consequence, as pointed out in Appendix~\ref{appendix: gate-dep stability}.

The main upshot is that both gate-dependence or more general contextual effects, together with non-Markovian noise, in \gls{rb} and almost surely in any other technique, would require a further dealing with the environment. One possible way forward could be to study non-Markovian gate-dependent \gls{rb} with Fourier analysis, similar to~\cite{merkel2018randomized}; this has the potential to be not only more compact and simple, but also to allow for more generality and to obtain deeper consequences.

\glsresetall 
\section{Conclusions}\label{sec: conclusions}
We have established a \gls{rb} framework for non-Markovian noise with gate sets forming a finite group which admits a multiplicity-free representation. Despite this being a more general and abstract case than that of unitary 2-designs~\cite{figueroaromero2021randomised}, it renders a much clearer functional form of the \gls{asf} as described in terms of quality maps, which we identify as the objects carrying average noise through the environment that mediates temporal correlations. Quality maps naturally generalize the concept of quality parameters~\cite{Helsen2019}, from Markovian to non-Markovian \gls{rb}, as the central quantities capturing average noise rates within the \gls{asf}. The main difference between quality maps and quality factors, is that the first capture all temporal correlations within a \gls{rb} experiment; as such, all information to be known about the intrinsic noise is carried through these, with all remaining quantities being experimental choices for the benchmarked gates, initial state and measurement.

The main obstacle to operationally extract average error rates from quality maps can be tracked back to the environment. Furthermore, the mere fact that prepared initial states and measurements can give rise to correlated \gls{spam} errors, appears as a downside to employing the \gls{rb} technique under this type of noise. Nevertheless, we have provided a means to bridge this gap and obtain an operational estimate of gate fidelities for full non-Markovian \gls{rb} processes by simply extending the \gls{rb} protocol to include a coherent averaging over initial states and measurements. This effectively removes the environmental functional dependence and renders a single figure of merit as an average fidelity of the full \gls{rb} process, including the correlated \gls{spam}. Further estimates of average noise of gates alone can then potentially be made through Character \gls{rb}~\cite{Helsen2019} or otherwise.

In non-Markovian \gls{rb}, given that noise rates at a given timestep will depend on all previous ones, gate fidelities for single gates make sense only when relative to a whole noise process. This prompts pointing out that purely non-Markovian behavior in \gls{rb} can be detected, and accounted for as above, whenever fidelities are seen to increase in subsequent timesteps. We obtained conditions under which these deviations can be observed, and in particular we highlight that it is necessary for the noise at a given step to increase the purity of the full $\syst\env$ state at such step by over $\dimS^2$, together with having an environment with a size over that of the system, $\dimE>\dimS$, to witness such increase in a subsequent step.

Finally, arguably one of the major hurdles to be cleared in any benchmarking or characterization technique, is incorporating non-Markovianity together with gate-dependence or a more general context-dependence. We noticed that, while the results of~\cite{Wallman_2018, Helsen2019}, showing that gate-dependence introduces a single perturbative term in the \gls{asf} (itself decaying exponentially in sequence length) do not extend trivially to the non-Markovian case, the stability bound of~\cite{Wallman_2014} on the variance of the sequence fidelity still holds in the non-Markovian case when incorporating the size of the environment, and similarly for the \gls{asf}.

An overall highlight in our analysis is that the role of the environment in non-Markovian \gls{rb} is to carry temporal correlations within the noise, as well as gate-dependence and potentially more general context dependence, through quality maps: further studying these quantities could give relevant insights into non-Markovian error mitigation or other advantageous uses of memory effects. This could be done, e.g., for specific microscopic models of noise stemming from the open quantum systems literature~\cite{RevModPhys.89.015001, RevModPhys.88.021002}, or with general approaches, such as treating \gls{rb} as convolution~\cite{merkel2018randomized}. While non-Markovianity has stayed relatively in the dark when considered in the context of noise characterization techniques, it almost certainly can be smoothly incorporated into a generalized framework for \gls{rb} in the spirit of~\cite{helsen2020general}, and further allow for a more comprehensive understanding and control of memory effects in quantum technologies.

\begin{acknowledgements}
The authors thank Joel Wallman for insightful discussion. KM acknowledges the support of Australian Research Council's Discovery Projects DP210100597 \& DP220101793, and the International Quantum U Tech Accelerator award by the US Air Force Research Laboratory.
\end{acknowledgements}

\appendix
\onecolumn

\section{Background preliminaries}
Here we present some of the essential background related to representation theory of finite groups as well as randomized benchmarking; for the first, references such as~\cite{tinkham2003group, harris1991representation} can be consulted for an in-depth treatment, while for \gls{rb} we mainly follow the results of~\cite{Helsen2019}.

\glsresetall
\subsection{Representations of finite groups}\label{appendix: representations of finite groups}
A group is a set $\mbb{G}$ and an operation $\cdot$ satisfying:
\begin{itemize}[nosep, label={$\diamond$}]
    \item (Closure) For all $g_1,g_2\in\mbb{G}$, also $g_1\cdot{g}_2=\mbb{G}$.
    \item (Associativity) For all $g_1,g_2,g_3\in\mbb{G}$, $(g_1\cdot{g}_2)\cdot{g}_3=g_1\cdot(g_2\cdot{g}_3)$.
    \item (Identity) There exists $e\in\mbb{G}$ such that $e\cdot{g}=g\cdot{e}=g$ for all $g\in\mbb{G}$.
    \item (Inverse) There exists $g^{-1}\in\mbb{G}$ for all $g\in\mbb{G}$ such that $g^{-1}\cdot{g}=g\cdot{g}^{-1}=e$.
\end{itemize}
Henceforth we omit the $\cdot$ symbol and take group multiplication to be the group operation.

A representation of a group provides a way of dealing with abstract groups as linear transformations to a vector space. In particular, we restrict ourselves to representations by unitary linear operators on $\mbb{U}(V)$ for a complex finite-dimensional vector space $V$, and we may define a representation of a group $\mbb{G}$ as a map
\begin{equation}
    \phi: \mbb{G}\to\mbb{U}(V):\,g\mapsto\phi(g),
\end{equation}
being such that
\begin{equation}
    \phi(g)\phi(h) = \phi(gh),\qquad\forall{g,h\in\mbb{G}},
\end{equation}
i.e., with the group operation being preserved under matrix multiplication of $\phi$. Here we will normally take $V$ as a vector space over $d\times{d}$ matrices.

We call a representation $\phi$ reducible if there exists some transformation $S$ such that
\begin{align}
    S\phi(g)S^{-1} &= \phi_1(g)\oplus\phi_2(g) = \begin{pmatrix}\phi_1(g)&0\\0&\phi_2(g)\end{pmatrix},
\end{align}
for all $g\in\mbb{G}$. Otherwise, a representation is called irreducible. This implies that any reducible representation can be written as a direct sum of irreducible ones: more generally, Maschke's theorem~\cite{harris1991representation} ensures that, for all $g\in\mbb{G}$
\begin{align}
    \phi(g) = \Moplus_{\pi\in{R}_{\mbb{G}}}\phi_\pi^{\otimes\mu_\pi}(g),
\end{align}
where $R_{\mbb{G}}$ is a set of labels for the irreducible representations, and $\mu_\pi$ is a non-negative integer denoting the multiplicity (equivalent copies) of $\phi_\pi$. In this manuscript we deal solely with groups admitting multiplicity-free representations, meaning $\mu_\pi=1$ for all $\pi$.

\subsection{Schur's lemma and twirling}\label{appendix: twirl}
A central result in representation theory is known as Schur's lemma: in a nutshell, it states that \emph{the only matrices that commute with all elements of an irreducible representation of a group are constant matrices (i.e., scalar multiples of $\mbb1$)}. This is stated as in Lemma 1 of~\cite{PhysRevLett.109.240504}, which in turn refers to~\cite{tinkham2003group} for detailed proofs; similarly other standard literature can be consulted for an in-depth treatment, e.g.,~\cite{harris1991representation}. Here we employ a consequence of Schur's lemma, rather than the lemma itself, applying to the so-called twirl of an operator given a representation of a group.

Consider $\mathbb{G}$ a finite group and $V$ be some finite dimensional complex vector space as above. For a representation $\phi$ of $\mathbb{G}$, the twirl $\mc{T}_\phi$ is defined by
\begin{equation}
    \mc{T}_\phi(A) := \f{1}{|\mathbb{G}|}\sum_{g\in\mathbb{G}} \phi(g)\,A\,\phi(g)^\dg,
\end{equation}
for all linear maps $A:V\to{V}$.

\begin{lemma}[\label{lemma: Schur}Lemma 1 in~\cite{Helsen2019}] Let $\mathbb{G}$ be a finite group and let $\phi$ be a multiplicity-free representation of $\mathbb{G}$ on a complex vector space $V$ with decomposition
\begin{equation}
    \phi(g) \simeq \Moplus_{\pi\in{R}_{\mathbb{G}}} \phi_\pi(g),\qquad\forall{g}\in\mathbb{G}
\end{equation}
into inequivalent irreducible subrepresentations $\phi_\pi$. Then for any linear map $A:V\to{V}$ the twirl of $A$ over $\mathbb{G}$ takes the form
\begin{equation}
    \mc{T}_\phi(A) = \sum_{\pi\in{R}_{\mathbb{G}}} \f{\tr(AP_\pi)}{\tr(P_\pi)}P_\pi,
\end{equation}
where $P_\pi$ is the projector onto the support of the representation $\phi_\pi$.
\end{lemma}

The multiplicity-free requirement is relaxed in~\cite{helsen2020general} in the context of Markovian randomized benchmarking, but here we will solely focus on the former case.

\subsection{The superoperator representation}\label{appendix: superoperator rep}
To apply Lemma~\ref{lemma: Schur}, we may use the so-called \emph{superoperator representation} (a.k.a. natural, Liouville or vectorized representation). The idea is to represent quantum channels as matrices acting on vectorized states (in an extended space of dimension squared). It is similar in spirit to the Choi-Jamio\l{}kowski representation~\cite{watrous2018theory}, in the sense that it maps channels to matrices, although different in that it does not necessarily maps them to a quantum state (density matrix). Here we are mainly going to need the definition
\begin{equation}
    \mathrm{vec}(|i\rangle\!\langle{j}|) := |ij\rangle,
    \label{eq: def vec map}
\end{equation}
so that generally
\begin{equation}
    |X\rrangle := \mathrm{vec}(X),
\end{equation} for any matrix $X$. This implies that for a {\gls{cp}} map $\Phi$,
\begin{equation}
    |\Phi(X)\rrangle := \hat{\Phi}|X\rrangle,
\end{equation}
where,
\begin{equation}
    \Phi(\cdot):=\sum_\mu\varphi_\mu(\cdot)\,\varphi_\mu^\dg,\,\quad\text{implies}\quad\,\hat{\Phi} = \sum_\mu \varphi_\mu \otimes \varphi_\mu^*,
\end{equation}
for Kraus operators $\varphi_\mu$ of $\Phi$, and where $*$ here means entry-wise conjugate\footnote{ This follows from ${\mathrm{vec}(AXB)=(A\otimes{B}^\mathrm{T})\mathrm{vec}(X)}$. Depending on the definition of the vec map, one may find this property with the right-hand side arranged differently. The definition in Eq.~\eqref{eq: def vec map} effectively stacks the rows of the matrix in a vector. A definition that is also common is ${\mathrm{vec}(|a\rangle\!\langle{b}|)=|ba\rangle}$, which in braket notation looks odd but when looking at matrices looks sort of natural because it stacks columns of the matrix into the vector; this definition would lead to a slightly different Liouville rep of quantum maps.}. The Liouville representation of $\Phi$ is here denoted by $\hat{\Phi}$, and we generally will distinguish maps, $\mc{X}$, when they are in the Liouville representation with a circumflex on top, $\hat{\mc{X}}$. The main point here is that quantum maps become matrices in this representation, which will let us work with them more easily without necessarily trying to extract properties of the maps in question. A clear exposition of this representation can be found e.g., in Section V.B of~\cite{Wallman_2014}.

\subsection{The standard randomized benchmarking protocol}\label{appendix: rb protocol}
A standard \gls{rb} protocol proceeds as follows:
\begin{enumerate}
    \item Prepare an initial state $\rho$ on the system of interest $\syst$.
    \item Sample $m$ distinct elements, $\mc{C}_1,\mc{C}_2,\ldots,\mc{C}_m$, uniformly at random from a given gate set $\mbb{K}$ containing the corresponding inverse elements. Let $\mc{C}_{m+1} := \Mcirc_{i=m}^1\mc{C}_i^{-1}=\mc{C}_1^{-1}\circ\cdots\circ\mc{C}_m^{-1}$, where $\circ$ denotes composition of maps. We refer to $\mc{C}_{m+1}$ as an undo-gate.
    \item Apply the composition $\Mcirc_{i=1}^{m+1}\mc{C}_i$ on $\rho$. In practice, this amounts to applying a noisy sequence $\mc{S}_m:=\Mcirc_{i=1}^{m+1}\mc{J}_i$ of length $m$ on $\rho$, where $\mc{J}_i$ are the physical noisy gates associated to $\mc{C}$.
    \item Estimate the probability $f_m := \tr\left[M{\mc{S}}_m\left(\rho\right)\right]$ via a \gls{povm} element $M$.
    \item Repeat $n$ times the steps 1 to 4 for the same initial state $\rho$, same \gls{povm}~element $M$, and different sets of gates chosen uniformly at random $\{\mc{C}_i^{(1)}\}_{i=1}^m,\{\mc{C}_i^{(2)}\}_{i=1}^m,\ldots,\{\mc{C}_i^{(n)}\}_{i=1}^m$ from $\mbb{K}$ to obtain the probabilities $f_m^{(1)},f_m^{(2)},\ldots{f}_m^{(n)}$. Compute the \gls{asf}, $\mc{F}_m=1/n\sum_{i=1}^nf_m^{(i)}$.
    \item Examine the behavior of the \gls{asf}, $\mc{F}_m$, over different sequence lengths $m$.
\end{enumerate}

\subsection{Markovian finite-group time-independent average sequence fidelity}
In standard, Markovian, time-independent and gate-independent \gls{rb}, we have a sequence
\begin{align}
    \mc{S}_m &:= \Mcirc_{i=1}^{m+1}\left(\Lambda\circ\mc{C}_i\right),
\end{align}
where $\mc{C}_{m+1}:= \mc{C}^\dg_1\circ\cdots\circ\mc{C}_m^\dg$ with $\mc{C}(\cdot)=G(\cdot)G^\dg$ and here $\mc{C}^\dg(\cdot)=G^\dg(\cdot)G$, both for $G\in\mathbb{G}$ unitary representations being the target gates and where the \gls{cptp}~map $\Lambda$ models the noise. Now, consider a change of variables $\mc{G}_j=\Mcirc_{i=1}^j\mc{C}_i$; this implies that, equivalently,
\begin{align}
    \mc{S}_m=\Lambda\circ\Mcirc_{i=1}^m\left(\mc{G}_i^\dg\circ\Lambda\circ\mc{G}_i\right).
\end{align}

Thus we have, for an initial state $\rho$ and a measurement with \gls{povm}~element $M$, over a sequence length $m$,
\begin{align}
    \mc{F}_m(\rho,M) &:= \tr\left\{{M}\,\mbb{E}\left[\mc{S}_m(\rho)\right]\right\} \nonumber\\
    &= \tr\left\{{M}\,\Lambda\circ\Mcirc_{i=1}^m\mbb{E}\left[\mc{G}_i^\dg\circ\Lambda\circ\mc{G}_i\right]\rho\right\}\nonumber\\
    &= \tr\left\{{M}\,\Lambda\circ\left(\mbb{E}\left[\mc{G}^\dg\circ\Lambda\circ\mc{G}\right]\right)^m\rho\right\},
    \label{eq: asf1}
\end{align}
where $\mbb{E}$ implicitly means uniform average over the gates $\mc{G}$. We will henceforth omit the explicit $\rho$ and $M$ dependence.

We can equivalently express Eq.~\eqref{eq: asf1} in the superoperator representation as
\begin{equation}
    \mc{F}_m = \llangle{M}\,|\hat{\Lambda}\left(\mbb{E}[\hat{\mc{G}}^\dg\hat{\Lambda}\hat{\mc{G}}]\right)^m|\,\rho\rrangle.
\end{equation}

Then it follows by Lemma~\ref{lemma: Schur}, that
\begin{align}
    \mc{F}_m &= \llangle{M}\,|\hat{\Lambda}\left(\sum_{\pi\in{R}_{\mathbb{G}}}f_\pi\hat{\mc{P}}_\pi\right)^m|\,\rho\rrangle \nonumber\\
    &= \sum_{\pi\in{R}_{\mathbb{G}}}f_\pi^m\,\llangle{M}\,|\hat{\Lambda}\hat{\mc{P}}_\pi|\,\rho\rrangle,
    \label{eq: asf wehner}
\end{align}
where here, as done in the main text and as we do onward, we denote by $\hat{\mc{P}}_\pi$ the projector onto the irreducible space corresponding to $\phi_\pi$, and where
\begin{equation}
    f_\pi := \f{\tr(\hat{\Lambda}\,\hat{\mc{P}}_\pi)}{\tr(\hat{\mc{P}}_\pi)},
\end{equation}
is called a \emph{quality parameter}, as in~\cite{Helsen2019}.

\subsection{Standard case: Markovian, time-independent, 2-design}\label{appendix: Markov 2 design}
Consider the case of the group $\mathbb{G}$ forming at least a 2-design, i.e., when uniformly averaging over it gives the same result as uniformly averaging over the whole unitary group. When we move to the superoperator representation, we may take the gates to the form $\hat{\mc{G}}=G\otimes{G}^*$ for unitary $G$ of dimension $\dimS$ and where $G^*$ denotes entry-wise complex conjugate; this is a representation with support on $\mbb{C}^{\dimS}\otimes\mbb{C}^{*\dimS}$, which can be decomposed into invariant subspaces having projectors (see e.g.~\cite{horodecki1997, chruscinski2006})
\begin{equation}
    \hat{\mc{P}}_1 = \Psi, \qquad
    \hat{\mc{P}}_2 = \mbb1 - \Psi,
\end{equation}
where $\Psi = \sum|ii\rangle\!\langle{jj}|/\dimS$ can be thought of as a partial transpose after a swap (and normalized). Hence too, assuming that $\Lambda$ is also trace-preserving,
\begin{align}
    f_1 &= \tr[\hat{\Lambda}{\Psi}]=\f{\sum\langle{i}|\lambda_\mu|j\rangle\!\langle{j}|\lambda_\mu^\dg|i\rangle}{\dimS} = 1,\\
    f_2 &= \f{\tr[\hat{\Lambda}]-\tr[\hat{\Lambda}{\Psi}]}{\dimS^2-1} = \f{\tr[\hat{\Lambda}]-1}{\dimS^2-1} := p,
\end{align}
where here $\lambda_\mu$ are the Kraus operators of the $\syst$ noise map $\Lambda$, and so as
\begin{align}
    \mc{F}_m = f_1^m \llangle{M}|\hat{\Lambda} \hat{\mc{P}}_1|\rho\rrangle + f_2^m \llangle{M}|\hat{\Lambda} \hat{\mc{P}}_2|\rho\rrangle,
\end{align}
and noticing that ${\Psi}|\rho\rrangle=\Psi\,\mathrm{vec}(\rho)=\sum|ii\rangle/\dimS=|\mbb1/\dimS\rrangle$, which is simply the superoperator representation of $\mbb1/\dimS$, this gives the standard result
\begin{align}
    \mc{F}_m &= \llangle{M}|\hat{\Lambda}|\f{\mbb1}{\dimS}\rrangle + p^m \llangle{M}|\hat{\Lambda}|\rho-\f{\mbb1}{\dimS}\rrangle = B + A p^m.
\end{align}

\section{Finite group non-Markovian randomized benchmarking}
\subsection{Time-independent non-Markovian noise}\label{appendix: nM finite group time-indep}
With this setup, the only difference for the case of non-Markovian \gls{rb}~is that now the noise $\Lambda$ acts jointly on a system $\syst$ and an environment $\env$, with respective dimensions $\dimS$ and $\dimE$, at every step of the \gls{rb}~sequence, and the gates $\mc{G}$ are now assumed to act solely on $\syst$~\cite{figueroaromero2021randomised}. Keep in mind that if we employ the superoperator representation, the environment will have implicit two copies\footnote{ We point out too that, strictly speaking, the order of the spaces would be grouped as $\mc{H}_\env\otimes\mc{H}_\syst\otimes\mc{H}_\env\otimes\mc{H}_\syst$. This is never a problem as long as we are consistent when acting on corresponding spaces: here we normally use $\mc{H}_\env\otimes\mc{H}_\env\otimes\mc{H}_\syst\otimes\mc{H}_\syst$.}. So we may now write Schur's lemma on subspace $\syst$ as
\begin{align}
    (\mc{I}_\env\otimes\mc{T}_\phi)\hat{\Lambda} &:= \f{1}{|\mathbb{G}|}\sum_{g\in\mathbb{G}}(\mbb1_\env\otimes\phi(g))\,\hat{\Lambda}\,(\mbb1_\env\otimes\phi(g)^\dg) \nonumber\\
    &= \sum_{e,\varepsilon,e^\prime,\varepsilon^\prime=1}^{\dimE}|ee^\prime\rangle\!\langle\varepsilon\varepsilon^\prime|\otimes\mc{T}_\phi\left(\hat{\Lambda}^{ee^\prime\varepsilon\varepsilon^\prime}_\syst\right)
    \nonumber\\
    &= \sum_{e,\varepsilon,e^\prime,\varepsilon^\prime=1}^{\dimE}|ee^\prime\rangle\!\langle\varepsilon\varepsilon^\prime|\otimes\sum_{\pi\in{R}_{\mathbb{G}}}f^{\,ee^\prime\varepsilon\varepsilon^\prime}_{\pi}\hat{\mc{P}}_\pi,
\end{align}
where $|e\rangle,|e^\prime\rangle,|\varepsilon\rangle,|\varepsilon^\prime\rangle$ are $\env$ orthonormal basis vectors and where now we define
\begin{align}
    f_\pi^{\,ee^\prime\varepsilon\varepsilon^\prime} &:= \f{\tr(\hat{\Lambda}^{ee^\prime\varepsilon\varepsilon^\prime}_\syst\,\hat{\mc{P}}_\pi)}{\tr(\hat{\mc{P}}_\pi)},\quad
    \text{with}\quad
    \hat{\Lambda}^{ee^\prime\varepsilon\varepsilon^\prime}_\syst = \sum_\mu \langle{e}|\lambda_\mu|\varepsilon\rangle \otimes \langle{e^\prime}|\lambda_\mu^*|\varepsilon^\prime\rangle, \label{eq: quality f env}
\end{align}
where here $\lambda_\mu$ are the Kraus operators of the \gls{cp} map $\Lambda$, which is acting on the full $\syst\env$ system (i.e., there is an implicit identity operator on $\syst$). Hence for the case of time-independent non-Markovian noise,
\begin{align}
    \mc{F}_m &= \llangle{M}\,|\hat{\tr}_\env\hat{\Lambda}\left(\mbb{E}[(\mc{I}_\env\otimes\hat{\mc{G}}^\dg)\hat{\Lambda}(\mc{I}_\env\otimes\hat{\mc{G}})]\right)^m|\,\rho\rrangle,\nonumber\\
    &= \llangle{M}\,|\hat{\tr}_\env\hat{\Lambda}\left(\sum_{e,e^\prime,\varepsilon,\varepsilon^\prime=1}^{\dimE}|ee^\prime\rangle\!\langle\varepsilon\varepsilon^\prime|\otimes\sum_{\pi\in{R}_{\mathbb{G}}}f^{\,ee^\prime\varepsilon\varepsilon^\prime}_{\pi}\hat{\mc{P}}_\pi\right)^m|\,\rho\rrangle,\nonumber\\
    &=\sum_{e,e^\prime,\varepsilon,\varepsilon^\prime=1}^{\dimE}\sum_{\pi\in{R}_{\mathbb{G}}}\boldsymbol{\mathbbm{f}}_{m,\pi}^{ee^\prime\varepsilon\varepsilon^\prime}\llangle{M}\,|\hat{\tr}_\env\hat{\Lambda}\left(|ee^\prime\rangle\!\langle\varepsilon\varepsilon^\prime|\otimes\hat{\mc{P}}_\pi\right)|\,\rho\rrangle,
    \label{eq: nM static ASF}
\end{align}
where
\begin{equation}
    \boldsymbol{\mathbbm{f}}_{m,\pi}^{ee^\prime\varepsilon\varepsilon^\prime} \! := \! \sum_{\{\epsilon_i=1,\epsilon_i^\prime=1\}_{i=1}^{m-1}}^{\dimE} f^{ee^\prime\epsilon_1\epsilon_1^\prime}_{\pi}f^{\epsilon_1\epsilon_1^\prime\epsilon_2\epsilon_2^\prime}_{\pi}f^{\epsilon_2\epsilon_2^\prime\epsilon_3\epsilon_3^\prime}_{\pi}\cdots f^{\epsilon_{m-1}\epsilon_{m-1}^\prime\varepsilon\varepsilon^\prime}_{\pi},
    \label{eq: quality tensor env}
\end{equation}
is now a quality factor containing all correlated individual quality factors defined in Eq.~\eqref{eq: quality f env}. Notice that a time-independence in the sense of all $\Lambda$ maps being identical at each timestep does not provide any simplification given the environmental dependence in each quality factor.

\subsection{Standard case: non-Markovian, time-independent, 2-design}\label{appendix: asf nM Clifford}
For the unitary 2-design case, as in Section~\ref{appendix: Markov 2 design}, we have $\hat{\mc{P}}_1=\Psi$ and $\hat{\mc{P}}_2=\mbb1-\Psi$ where here again $\Psi=\sum|ii\rangle\!\langle{jj}|/\dimS$. Now, however,
\begin{align}
    f_1^{ee^\prime\varepsilon\varepsilon^\prime} &= \tr[\hat{\Lambda}_\mathsf{S}^{ee^\prime\varepsilon\varepsilon^\prime}\Psi] \nonumber\\
    &= \f{\sum\langle{e}i|\lambda_\mu|\varepsilon{j}\rangle\!\langle{e^\prime}i|\lambda_\mu^*|\varepsilon^\prime{j}\rangle}{\dimS} \nonumber\\
    &= \f{\sum\langle{e}|\tr_\syst\left[\lambda_\mu|\varepsilon\rangle\!\langle\varepsilon^\prime|\lambda_\mu^\dg\right]|e^\prime\rangle}{\dimS} \nonumber\\
    &= \langle{e}|\Theta_\Lambda(|\varepsilon\rangle\!\langle\varepsilon^\prime|)|e^\prime\rangle, \\
    f_2^{ee^\prime\varepsilon\varepsilon^\prime} &= \f{\tr[\hat{\Lambda}_\mathsf{S}^{ee^\prime\varepsilon\varepsilon^\prime}]-\tr[\hat{\Lambda}_\mathsf{S}^{ee^\prime\varepsilon\varepsilon^\prime}\Psi]}{\dimS^2-1} \nonumber\\
    &= \f{\sum \langle{e}|\tr_\syst(\lambda_\mu)|\varepsilon\rangle\!\langle{e}^\prime|\tr_\syst(\lambda_\mu^*)|\varepsilon^\prime\rangle-\langle{e}|\Theta(|\varepsilon\rangle\!\langle\varepsilon^\prime|)|e^\prime\rangle}{\dimS^2-1} \nonumber\\
    &= \f{\sum \langle{e}|\tr_\syst(\lambda_\mu)|\varepsilon\rangle\!\langle\varepsilon^\prime|\tr_\syst(\lambda_\mu^\dg)|e^\prime\rangle-\langle{e}|\Theta(|\varepsilon\rangle\!\langle\varepsilon^\prime|)|e^\prime\rangle}{\dimS^2-1} \nonumber\\
    &= \langle{e}|\,\f{(\$_\Lambda-\Theta_\Lambda)(|\varepsilon\rangle\!\langle\varepsilon^\prime|)}{\dimS^2-1}\,|e^\prime\rangle,
\end{align}
where here $\lambda_\mu$ are Kraus operators of the full $\syst\env$ noise map $\Lambda$, and the maps $\$_\Lambda$ and $\Theta_\Lambda$ are exactly those defined in~\cite{figueroaromero2021randomised},
\begin{align}
    \Theta_\Lambda(\cdot) &= \tr_\syst\left[\Lambda\left(\cdot\otimes\f{\mbb1}{\dimS}\right)\right],
    \label{eq: dollar map}\\
    \$_\Lambda(\cdot) &:= \sum_\mu\tr_\syst(\lambda_\mu)(\cdot)\tr_\syst(\lambda_\mu^\dg).
    \label{eq: theta map}
\end{align}

The compositions of the $\$$ and $\Theta$ maps are now simply products: first, we have for $m=2$,
\begin{align}
    \mathbbm{f}_{m=2,\pi=1}^{ee^\prime\varepsilon\varepsilon^\prime} &= \sum_{\epsilon,\epsilon^{\prime}}f_1^{ee^\prime\epsilon\epsilon^\prime}f_1^{\epsilon\epsilon^\prime\varepsilon\varepsilon^\prime} \nonumber\\
    &= \sum_{\epsilon,\epsilon^{\prime}} \langle{e}|\Theta_\Lambda(|\epsilon\rangle\!\langle\epsilon^\prime|)|e^\prime\rangle\!\langle{\epsilon}|\Theta_\Lambda(|\varepsilon\rangle\!\langle\varepsilon^\prime|)|\epsilon^\prime\rangle \nonumber\\
    &= \sum_{\epsilon,\epsilon^{\prime}} \langle{e}|\Theta_\Lambda\left\{|\epsilon\rangle\!\langle{\epsilon}|\Theta_\Lambda(|\varepsilon\rangle\!\langle\varepsilon^\prime|)|\epsilon^\prime\rangle\!\langle\epsilon^\prime|\right\}|e^\prime\rangle \nonumber\\
    &= \langle{e}|\Theta_\Lambda^{\circ2}(|\varepsilon\rangle\!\langle\varepsilon^\prime|)|e^\prime\rangle,
\end{align}
so that it follows that $\mathbbm{f}_{m,\pi=1}^{ee^\prime\varepsilon\varepsilon^\prime}=\langle{e}|\Theta_\Lambda^{\circ{m}}(|\varepsilon\rangle\!\langle\varepsilon^\prime|)|e^\prime\rangle$. Similarly,
\begin{align}
    \mathbbm{f}_{m=2,\pi=2}^{ee^\prime\varepsilon\varepsilon^\prime} &= \sum_{\epsilon,\epsilon^{\prime}}f_2^{ee^\prime\epsilon\epsilon^\prime}f_2^{\epsilon\epsilon^\prime\varepsilon\varepsilon^\prime} \nonumber\\
    &= \f{1}{(\dimS^2-1)^2}\sum_{\epsilon,\epsilon^{\prime}} \langle{e}|(\$_\Lambda-\Theta_\Lambda)(|\epsilon\rangle\!\langle\epsilon^\prime|)|e^\prime\rangle\!\langle\epsilon|(\$_\Lambda-\Theta_\Lambda)(|\varepsilon\rangle\!\langle\varepsilon^\prime|)|\epsilon^\prime\rangle \nonumber\\
    &= \f{\langle{e}|(\$_\Lambda-\Theta_\Lambda)^{\circ2}(|\varepsilon\rangle\!\langle\varepsilon^\prime|)|e^\prime\rangle}{(\dimS^2-1)^2},
\end{align}
so that $\mathbbm{f}_{m,\pi=2}^{ee^\prime\varepsilon\varepsilon^\prime}= (\dimS^2-1)^m\langle{e}|(\$_\Lambda-\Theta_\Lambda)^{\circ{m}}(|\varepsilon\rangle\!\langle\varepsilon^\prime|)|e^\prime\rangle$.

For the remaining quantities we have
\begin{align}
    (|ee^\prime\rangle\!\langle\varepsilon\varepsilon^\prime|\otimes\Psi)\,|\rho\rrangle &= (|ee^\prime\rangle\!\langle\varepsilon\varepsilon^\prime|\otimes\Psi)\,\mathrm{vec}(\rho) \nonumber\\
    &= \f{1}{\dimS}\sum\rho_{e_0e_0^\prime{s}_0s_0^\prime}|ee^\prime{ii}\rangle\!\langle\varepsilon\varepsilon^\prime{jj}|e_0e_0^\prime{s}_0s_0^\prime\rangle \nonumber\\
    &=\f{1}{\dimS}\sum\rho_{e_0e_0^\prime{s}_0s_0^\prime}|ee^\prime{ii}\rangle\!\langle\varepsilon\varepsilon^\prime{jj}|e_0e_0^\prime{s}_0s_0^\prime\rangle \nonumber\\
    &= \f{1}{\dimS}\sum\rho_{\varepsilon\varepsilon^\prime{s}_0s_0}|ee^\prime{ii}\rangle \nonumber\\
    &= \langle\varepsilon|\rho_\env|\varepsilon^\prime\rangle\,|\left(|e\rangle\!\langle{e}^\prime|\otimes\f{\mbb1}{\dimS}\right)\rrangle,
    \end{align}
and thus similarly,
\begin{align}
    (|ee^\prime\rangle\!\langle\varepsilon\varepsilon^\prime|\otimes\mbb1)\,|\rho\rrangle &= \sum\rho_{\varepsilon\varepsilon^\prime{ij}}|ee^\prime{ij} \rangle\nonumber\\
    &= \mathrm{vec}\left(\rho_{\varepsilon\varepsilon^\prime{ij}}|e\rangle\!\langle{e}^\prime|\otimes|i\rangle\!\langle{j}|\right)\nonumber\\
    &= |\left(|e\rangle\!\langle{e}^\prime|\otimes\langle\varepsilon|\rho|\varepsilon^\prime\rangle\right)\rrangle,
\end{align}
which implies that
\begin{align}
    (|ee^\prime\rangle\!\langle\varepsilon\varepsilon^\prime|\otimes\hat{\mc{P}}_2)\,|\rho\rrangle &= |\left(|e\rangle\!\langle{e}^\prime|\otimes\langle\varepsilon|\rho|\varepsilon^\prime\rangle\right)\rrangle - \langle\varepsilon|\rho_\env|\varepsilon^\prime\rangle\,|\left(|e\rangle\!\langle{e}^\prime|\otimes\f{\mbb1}{\dimS}\right)\rrangle.
\end{align}

To finally put together the whole \gls{asf}, we have
\begin{align}
    \sum_{e,e^\prime,\varepsilon,\varepsilon^\prime}\mathbbm{f}_{m,\pi=1}^{ee^\prime\varepsilon\varepsilon^\prime} \llangle{M}|\hat{\tr}_\env\hat{\Lambda}(|ee^\prime\rangle\!\langle\varepsilon\varepsilon^\prime|\otimes\hat{\mc{P}}_1)|\,\rho\rrangle &= \sum \langle{e}|\Theta_\Lambda^{\circ{m}}(\rho_\env)|e^\prime\rangle\,\llangle{M}|\hat{\tr}_\env\hat{\Lambda}\,|\left(|e\rangle\!\langle{e}^\prime|\otimes\f{\mbb1}{\dimS}\right)\rrangle \nonumber\\
    &= \llangle{M}|\hat{\tr}_\env\hat{\Lambda}\,|\left(\Theta_\Lambda^{\circ{m}}(\rho_\env)\otimes\f{\mbb1}{\dimS}\right)\rrangle \nonumber\\
    &:= \llangle{M}|\hat{\tr}_\env\hat{\Lambda}\,|\mscr{B}_m(\rho)\rrangle
\end{align}
where we identified $\mscr{B}_m(\rho) := \Theta_\Lambda^{\circ{m}}(\rho_\env)\otimes\f{\mbb1}{\dimS}$ from~\cite{figueroaromero2021randomised} and
\begin{align}
    \sum_{e,e^\prime,\varepsilon,\varepsilon^\prime}\mathbbm{f}_{m,\pi=2}^{ee^\prime\varepsilon\varepsilon^\prime} \llangle{M}|\hat{\tr}_\env\hat{\Lambda}(|ee^\prime\rangle\!\langle\varepsilon\varepsilon^\prime|\otimes\hat{\mc{P}}_2)|\,\rho\rrangle &= \sum \f{\llangle{M}|\hat{\tr}_\env\hat{\Lambda}\,|\left[(\$_\Lambda-\Theta_\Lambda)^{\circ{m}}(|\varepsilon\rangle\!\langle\varepsilon^\prime|)\otimes\langle\varepsilon|\rho|\varepsilon^\prime\rangle\right]\rrangle}{(\dimS^2-1)^m} \nonumber\\
    &\qquad-\f{\llangle{M}|\hat{\tr}_\env\hat{\Lambda}\,|\left[(\$_\Lambda-\Theta_\Lambda)^{\circ{m}}(\rho_\env)\otimes\f{\mbb1}{\dimS}\right]\rrangle}{(\dimS^2-1)^m} \nonumber\\
    &= \f{\llangle{M}|\hat{\tr}_\env\hat{\Lambda}\,|\left[(\$_\Lambda-\Theta_\Lambda)^{\circ{m}}\otimes\mc{I}_\syst\right](\rho-\rho_\env\otimes\f{\mbb1}{\dimS})\rrangle}{(\dimS^2-1)^m} \nonumber\\
    &:= \llangle{M}|\hat{\tr}_\env\hat{\Lambda}\,|\mscr{A}_m(\rho)\rrangle,
\end{align}
where $\mscr{A}_m(\rho) := \left[(\$_\Lambda-\Theta_\Lambda)^{\circ{m}}\otimes\mc{I}_\syst\right](\rho-\rho_\env\otimes\f{\mbb1}{\dimS})$ as also defined in~\cite{figueroaromero2021randomised}, so that indeed we recover the full expression
\begin{equation}
    \mc{F}_m = \llangle{M}|\hat{\tr}_\env\hat{\Lambda}\,|\mscr{A}_m+\mscr{B}_m(\rho)\rrangle.
\end{equation}

\subsection{From quality parameters to quality maps}
In general, we can think of the whole term
\begin{align}
    \hat{\mc{Q}}_{m,\pi} := \sum_{e,e^\prime,\varepsilon,\varepsilon^\prime}\mathbbm{f}_{m,\pi}^{ee^\prime\varepsilon\varepsilon^\prime}|ee^\prime\rangle\!\langle\varepsilon\varepsilon^\prime|,    
\end{align}
where $\mathbbm{f}_{m,\pi}^{ee^\prime\varepsilon\varepsilon^\prime}$ is the quality factor defined in Eq.~\eqref{eq: quality tensor env}, as the Liouville representation of an object that we may label as a \emph{quality map},
\begin{equation}
    \mc{Q}_{m,\pi}(\cdot) := \sum_{e,e^\prime,\varepsilon,\varepsilon^\prime}\mathbbm{f}_{m,\pi}^{ee^\prime\varepsilon\varepsilon^\prime}|e\rangle\!\langle\varepsilon|(\cdot)|\varepsilon^\prime\rangle\!\langle{e}^\prime|,
    \label{eq: quality map}
\end{equation}
or equivalently we can label this object generally as a \emph{quality tensor}, given that each $\mathbbm{f}_{m,\pi}$ can be thought of precisely as an \gls{mpo}~with quality parameter nodes and environment bonds.

We may then equivalently write the full non-Markovian \gls{asf}~as
\begin{align}
    \mc{F}_m &= \sum_{\pi\in{R}_{\mathbb{G}}} \llangle{M}\,|\hat{\tr}_\env\hat{\Lambda}(\hat{\mc{Q}}_{m,\pi}\otimes\hat{\mc{P}}_\pi)\,|\,\rho\rrangle \nonumber\\
    &= \sum_{\pi\in{R}_{\mathbb{G}}}\tr[{M}\,\tr_\env\circ\Lambda\circ(\mc{Q}_{m,\pi}\otimes\mc{P}_\pi)\,\rho],
    \label{eq: non-Markov asf}
\end{align}
with $\mc{P}_\pi$ the map associated to the projector $\hat{\mc{P}}_\pi$: these will depend on each particular group case and in general they are not projective maps (as one would initially be led to think), e.g., for the unitary 2-design case $\hat{\mc{P}}_1=\Psi$ corresponds to $\mc{P}_1(\cdot)=\f{\mbb1}{\dimS}\tr(\cdot)$ and $\hat{\mc{P}}_2=\mbb1-\Psi$ corresponds to $\mc{P}_2(\cdot)=(\cdot)-\f{\mbb1}{\dimS}\tr(\cdot)$.

\subsection{Time-dependent noise}\label{appendix: time-dependent noise}
We can generalize the previous \gls{asf}~to the case of time-dependent noise in a straightforward way. For the Markovian case,
\begin{align}
    \mc{F}_m^\markov &= \llangle{M}|\hat{\Lambda}_{m+1}\,\prod_{i=1}^m\mbb{E}[\hat{\mc{G}}_i^\dg\hat{\Lambda}_i\hat{\mc{G}}_i]|\,\rho\rrangle \nonumber\\
    &= \sum_{\pi\in{R}_{\mathbb{G}}}\mathbbm{f}_{m,\pi}^\markov\llangle{M}|\hat{\Lambda}_{m+1}\hat{\mc{P}}_\pi|\,\rho\rrangle,
\end{align}
where here $\Lambda$ acts solely on $\syst$ and where
\begin{equation}
    \mathbbm{f}_{m,\pi}^\markov:=f_{(1),\pi}f_{(2),\pi} \cdots f_{(m),\pi}\quad \text{with}\quad f_{(i),\pi} := \f{\tr(\hat{\Lambda}_i\hat{\mc{P}}_\pi)}{\tr(\hat{\mc{P}}_\pi)}.
    \label{eq: time-dep Markov quality}
\end{equation}

Similarly, for the non-Markovian case, now with $\Lambda$ acting on the full $\syst\env$,
\begin{align}
    \mc{F}_m &=\!\!\sum_{e,e^\prime,\varepsilon,\varepsilon^\prime=1}^{\dimE}\sum_{\pi\in{R}_{\mathbb{G}}}\mathbbm{f}_{m,\pi}^{ee^\prime\varepsilon\varepsilon^\prime}\llangle{M}|\hat{\tr}_\env\hat{\Lambda}_{m+1}\!\left(|ee^\prime\rangle\!\langle\varepsilon\varepsilon^\prime|\otimes\hat{\mc{P}}_\pi\right)|\,\rho\rrangle,
\end{align}
where now
\begin{equation}
    \mathbbm{f}_{m,\pi}^{ee^\prime\varepsilon\varepsilon^\prime} := \sum_{\{\epsilon_i,\epsilon_i^\prime\}}f^{ee^\prime\epsilon_1\epsilon_1^\prime}_{(1),\pi}f^{\epsilon_1\epsilon_1^\prime\epsilon_2\epsilon_2^\prime}_{(2),\pi}\cdots{f}^{\epsilon_{m-2}\epsilon_{m-2}^\prime\epsilon_{m-1}\epsilon_{m-1}^\prime}_{(m-1),\pi}f^{\epsilon_{m-1}\epsilon_{m-1}^\prime\varepsilon\varepsilon^\prime}_{(m),\pi},
    \label{eq: time-dep nM quality}
\end{equation}
with
\begin{equation}
    f_{(i),\pi}^{\,ee^\prime\varepsilon\varepsilon^\prime} := \f{\tr(\hat{\Lambda}^{ee^\prime\varepsilon\varepsilon^\prime}_{i,\syst}\,\hat{\mc{P}}_\pi)}{\tr(\hat{\mc{P}}_\pi)},\quad\text{where}\quad
    \hat{\Lambda}^{ee^\prime\varepsilon\varepsilon^\prime}_{i,\syst} = \sum_\mu \langle{e}|\lambda_\mu^{(i)}|\varepsilon\rangle \otimes \langle{e^\prime}|\lambda_\mu^{(i)\,*}|\varepsilon^\prime\rangle,
\end{equation}
where here $\lambda_\mu^{(i)}$ are the Kraus operators of the respective $\syst\env$ map $\Lambda_i$.

Notice that non-Markovianity is a time-dependence in itself, with the time-labels constituting an extra layer of time-dependence where the noise maps $\Lambda$ can themselves differ arbitrarily between timesteps. The crucial difference between Markovian and non-Markovian time-dependence, however, is that the Markovian one simply refers to an explicit dependence on timesteps but not to a temporal correlation between these: this is clear if one compares Eq.~\eqref{eq: time-dep Markov quality} with its non-Markovian counterpart in Eq.~\eqref{eq: time-dep nM quality}: it is apparent that Markovian time-dependent {\gls{asf}}s can be reproduced by non-Markovian time-independent ones with a large enough environment.

\subsection{The quality map in the Markov limit}\label{appendix: quality Markov limit}
Consider a single step, $m=1$, and two distinct \gls{asf}, one Markovian and the other non-Markovian. Suppose for the Markov case we have some \gls{cp}~map $\Phi$ acting on $\syst$ such that
\begin{align}
    \mathbbm{f}_{m=1,\pi}^\markov = f_{(1),\pi} = \f{\tr[\hat{\Phi} \hat{\mc{P}}_\pi]}{\tr[\hat{\mc{P}}_\pi]} = \sum_
    \mu\f{\tr[(\phi_\mu\otimes\phi_\mu^*)\hat{\mc{P}}_\pi]}{\tr[\hat{\mc{P}}_\pi]},
\end{align}
while for some other non-Markovian case we have an $\syst\env$ map $\Lambda$ such that
\begin{align}
    \mathbbm{f}_{m=1, \pi}^{ee^\prime\varepsilon\varepsilon^\prime} =  f_{(1),\pi}^{ee^\prime\varepsilon\varepsilon^\prime} &= \f{\tr[\hat{\Lambda}_{\syst}^{ee^\prime\varepsilon\varepsilon^\prime}\hat{\mc{P}}_\pi]}{\tr[\hat{\mc{P}}_\pi]} \nonumber\\
    &= \sum_\mu \f{\tr[(\langle{e}|\lambda_\mu^{(i)}|\varepsilon\rangle \otimes \langle{e^\prime}|\lambda_\mu^{(i)\,*}|\varepsilon^\prime\rangle)\hat{\mc{P}}_\pi]}{\tr[\hat{\mc{P}}_\pi]}.
\end{align}

The two quantities are actually somewhat akin: let $\Phi$ have a Stinespring dilation representation ${\Phi(\cdot) = \tr_\mc{E}[\mc{U}(\nu\otimes\cdot)]}$ for some superfluous environment $\mc{E}$, an $\syst\mc{E}$ unitary channel $\mc{U}$ and a pure state $\nu:=|\nu\rangle\!\langle\nu|$ on $\mc{E}$. Then the Kraus operators of $\Phi$ are $\phi_\mu=\langle\mu|\mc{U}|\nu\rangle$, so that
\begin{equation}
    \mathbbm{f}_{m=1,\pi}^\markov = \sum_\mu\f{\tr[(\langle\mu|\mc{U}|\nu\rangle\otimes\langle\mu|\mc{U}^*|\nu\rangle)\hat{\mc{P}}_\pi]}{\tr[\hat{\mc{P}}_\pi]}.
    \label{eq: Markov quality param dilation}
\end{equation}

The non-Markovian quality parameter will reduce to something of this form in the Markov limit: indeed, in such a case, within the full \gls{asf}, we can interchange the partial trace with the noise due to the undo gate, which would have the form $\Lambda_2=\mc{I}_\env\otimes\Lambda_2^\markov$, thus rendering a term $\delta_{ee^\prime}$, and hence
\begin{align}
    \mathbbm{f}_{m=1, \pi}^{ee^\prime\varepsilon\varepsilon^\prime} \stackrel{\text{Markov}}{\longrightarrow} \sum_{e}\sum_\mu \f{\tr[(\langle{e}|\lambda_\mu^{(i)}|\varepsilon\rangle \otimes \langle{e}|\lambda_\mu^{(i)\,*}|\varepsilon^\prime\rangle)\hat{\mc{P}}_\pi]}{\tr[\hat{\mc{P}}_\pi]},
\end{align}
reducing effectively to Eq.~\eqref{eq: Markov quality param dilation}. For the general sequence length case, this limit essentially corresponds to \emph{Markovianizing} the process tensor as described in~\cite{figueroaromero2021randomised}, i.e., tracing out the environment at every step.

\subsection{Average sequence fidelity finite memory perturbation}\label{appendix: finite memory asf}
Consider now noise maps that take a convex combination form,
\begin{equation}
    \Gamma_i := q_i\,\Lambda_i + (1-q_i)\,(\mc{I}_\env\otimes\Phi_i), \quad \text{with} \quad 0\leq q_i \leq1,
\end{equation}
where $\Phi_i$ acts solely on system $\syst$ while $\Lambda_i$ acts on $\syst\env$, and we are interested in the case where $q_i$ is small and much less than $1/2$. Because the Kraus operators of this map are of the form $\{\sqrt{q_i}\lambda_\mu^{(i)},\,\sqrt{1-q_i}\,\phi_\nu^{(i)}\}$, the quality parameters are
\begin{equation}
    f_{(i),\pi}^{ee^\prime\varepsilon\varepsilon^\prime}(\hat{\Lambda}_i) = q_i f_{(i),\pi}^{ee^\prime\varepsilon\varepsilon^\prime}(\Lambda_i) + (1-q_i)\,\delta_{e\varepsilon}\delta_{e^\prime\varepsilon^\prime}\,f_{(i),\pi}(\Phi_i),
\end{equation}
where the dependence on the noise maps is explicit.

Thus, consider first the case $m=2$; we will now omit the noise map dependence, as it should be clear whether these refer for $\Lambda_i$ or $\Phi_i$ whenever they have upper $\env$ indices or otherwise, let us also omit the $\pi$ subindices for now and let $r_i:=(1-q_i)$ to shorten notation, then
\begin{align}
    \boldsymbol{\mathbbm{f}}_{m=2}^{ee^\prime\varepsilon\varepsilon^\prime}\left(\undervec{\Gamma}\right) &:= \sum_{\epsilon,\epsilon^\prime=1}^{\dimE} f^{ee^\prime\epsilon\epsilon^\prime}_{(1)}(\Gamma_1)f^{\epsilon\epsilon^\prime\varepsilon\varepsilon^\prime}_{(2)}(\Gamma_2) \nonumber\\
    &=\sum_{\epsilon,\epsilon^\prime=1}^{\dimE}q_1q_2 f_{(1)}^{ee^\prime\epsilon\epsilon^\prime}f_{(2)}^{\epsilon\epsilon^\prime\varepsilon\varepsilon^\prime} + \sum_{\epsilon,\epsilon^\prime=1}^{\dimE} r_1r_2\,\delta_{e\epsilon}\delta_{e^\prime\epsilon^\prime}\delta_{\epsilon\varepsilon}\delta_{\epsilon^\prime\varepsilon^\prime}\,f_{(1)}f_{(2)}\nonumber\\
    &\qquad+ \sum_{\epsilon,\epsilon^\prime=1}^{\dimE}\left[(1-q_1)q_2\delta_{\epsilon\varepsilon}\delta_{\epsilon^\prime\varepsilon^\prime}f_{(1)}^{ee^\prime\epsilon\epsilon^\prime}f_{(2)} + q_1r_2\delta_{e\epsilon}\delta_{e^\prime\epsilon^\prime}f_{(1)}f_{(2)}^{\epsilon\epsilon^\prime\varepsilon\varepsilon^\prime}\right] \nonumber \\
    &= r_1r_2\delta_{e\varepsilon}\delta_{e^\prime\varepsilon^\prime} \mathbbm{f}_{m=2}^{\markov}\left(\undervec{\Phi}\right) + r_1q_2\,f_{(1)}^{ee^\prime\varepsilon\varepsilon^\prime}f_{(2)} + q_1r_2\,f_{(1)}f_{(2)}^{ee^\prime\varepsilon\varepsilon^\prime} + q_1q_2\,\boldsymbol{\mathbbm{f}}_{m=2}^{ee^\prime\varepsilon\varepsilon^\prime}\left(\undervec{\Lambda}\right),
\end{align}
where $\undervec{\Gamma}$ here denotes $\Gamma_i,\Gamma_j,\ldots,\Gamma_m$ with $i<j<m$ and similarly for $\undervec{\Lambda}$ and $\undervec{\Phi}$.

If the noise is time-independent then the general $m$ case will simply be a binomial expansion. Otherwise we may express the non-Markov contribution as a sum over permutations of Markov and non-Markov quality parameters. Let us look then at $m=3$; again keep in mind that $r_i:=(1-q_i)$, then
    \begin{align}
     &\boldsymbol{\mathbbm{f}}_{m=3,\pi}^{ee^\prime\varepsilon\varepsilon^\prime}\left(\undervec{\Gamma}\right) := \sum_{\{\epsilon_{i}^{(\prime)}\}} f^{ee^\prime\epsilon_1\epsilon_1^\prime}_{(1)}(\Gamma_1)f^{\epsilon_1\epsilon_1^\prime\epsilon_2\epsilon_2^\prime}_{(2)}(\Gamma_2)f^{\epsilon_2\epsilon_2^\prime\varepsilon\varepsilon^\prime}_{(3)}(\Gamma_3) \nonumber\\[0.1in]
    &= \prod_{n=1}^3r_n\delta_{e\varepsilon}\delta_{e^\prime\varepsilon^\prime} \mathbbm{f}_{m=3}^{\markov}\left(\undervec{\Phi}\right) \nonumber \\
    &+ \left[q_1r_2r_3\mathbbm{f}_{m=2}^\markov(\Phi_2,\Phi_3)\mathbbm{f}_{m=1}^{ee^\prime\varepsilon\varepsilon^\prime}\!(\Lambda_1) + r_1q_2r_3\mathbbm{f}_{m=2}^\markov(\Phi_1,\Phi_3)\mathbbm{f}_{m=1}^{ee^\prime\varepsilon\varepsilon^\prime}\!(\Lambda_2) + r_1r_2q_3\mathbbm{f}_{m=2}^\markov(\Phi_1,\Phi_2)\mathbbm{f}_{m=1}^{ee^\prime\varepsilon\varepsilon^\prime}\!(\Lambda_3)\right] \nonumber\\[0.1in]
    &+ \left[r_1q_2q_3\mathbbm{f}_{m=1}^\markov(\Phi_1)\mathbbm{f}_{m=2}^{ee^\prime\varepsilon\varepsilon^\prime}(\Lambda_2,\Lambda_3) + q_1r_2q_3\mathbbm{f}_{m=2}^{ee^\prime\varepsilon\varepsilon^\prime}(\Lambda_1,\Lambda_3)\mathbbm{f}_{m=1}^\markov(\Phi_2) + q_1q_2r_3\mathbbm{f}_{m=2}^{ee^\prime\varepsilon\varepsilon^\prime}(\Lambda_1,\Lambda_2)\mathbbm{f}_{m=1}^\markov(\Phi_3)\right] \nonumber\\
    &+ \prod_{n=1}^3q_n\,\boldsymbol{\mathbbm{f}}_{m=3}^{ee^\prime\varepsilon\varepsilon^\prime}\left(\undervec{\Lambda}\right) \nonumber\\
    &=  \prod_{n=1}^3(1-q_n)\delta_{e\varepsilon}\delta_{e^\prime\varepsilon^\prime} \mathbbm{f}_{m=3}^{\markov}\left(\undervec{\Phi}\right) + 
    \binom{(1-q)^2q\,\mathbbm{f}_{m=2}^{\markov}\mathbbm{f}_{m=1}^{ee^\prime\varepsilon\varepsilon^\prime}}{\mathbbm{f}_{m=1}^{ee^\prime\varepsilon\varepsilon^\prime}} + \binom{(1-q)q^2\mathbbm{f}_{m=1}^{\markov}\mathbbm{f}_{m=2}^{ee^\prime\varepsilon\varepsilon^\prime}}{\mathbbm{f}_{m=2}^{ee^\prime\varepsilon\varepsilon^\prime}} \nonumber\\
    &\qquad\qquad+ \prod_{n=1}^3q_n\,\boldsymbol{\mathbbm{f}}_{m=3}^{ee^\prime\varepsilon\varepsilon^\prime}\left(\undervec{\Lambda}\right),
\end{align}
where the notation $\displaystyle{\binom{(1-q)^\ell q^{m-\ell}\mathbbm{f}_{\ell}^{\markov}\mathbbm{f}_{m-\ell}^{ee^\prime\varepsilon\varepsilon^\prime}}{\mathbbm{f}_{m-\ell}^{ee^\prime\varepsilon\varepsilon^\prime}}}$ for $\ell\leq{m}$ here denotes the sum of possible combinations of the map arguments $\{q_i\Lambda_i\}$ in $\mathbbm{f}_{m-\ell}^{ee^\prime\varepsilon\varepsilon^\prime}$ within a product $\mathbbm{f}_{\ell}^{\markov}\mathbbm{f}_{m-\ell}^{ee^\prime\varepsilon\varepsilon^\prime}$, conditioned with the map arguments to be time-ordered, i.e., $(1-q_{i_1})\cdots(1-q_{i_\ell})q_{\ell+1}\cdots{q}_{i_m}\mathbbm{f}_\ell^{\markov}(\Phi_{i_1},\ldots,\Phi_{i_\ell})\mathbbm{f}_{m-\ell}^{ee^\prime\varepsilon\varepsilon^\prime}(\Lambda_{i_{\ell+1}},\ldots,\Lambda_{i_m})$ must be such that $i_1<\cdots<i_\ell<i_{\ell+1}<\cdots<i_m$. This is irrelevant in the Markovian quality parameters and could be relaxed but an extra factor would need to be added to avoid over-counting. The notation is suggestive because in the static case these will just correspond to binomial coefficients and powers of the Markovian quality parameters as $\binom{m}{\ell}(1-q)^\ell q^{m-\ell}f^{\ell}\mathbbm{f}_{m-\ell}^{ee^\prime\varepsilon\varepsilon^\prime}$.

We may thus write the general $m$ case as,
\begin{align}
     &\boldsymbol{\mathbbm{f}}_{m,\pi}^{ee^\prime\varepsilon\varepsilon^\prime}\left(\undervec{\Gamma}\right) = \prod_{n=1}^m(1-q_n)\delta_{e\varepsilon}\delta_{e^\prime\varepsilon^\prime}\mathbbm{f}_{m,\pi}^\markov(\undervec{\Phi}) + \sum_{\ell=0}^{m-1}\binom{q^{m-\ell}(1-q)^{\ell}\mathbbm{f}_{\ell,\,\pi}^{\markov}\mathbbm{f}_{m-\ell,\pi}^{ee^\prime\varepsilon\varepsilon^\prime}}{\mathbbm{f}_{m-\ell,\pi}^{ee^\prime\varepsilon\varepsilon^\prime}},
\end{align}
which in turn implies
\begin{align}
    \hat{\mc{Q}}_{m,\pi} &= \sum_{e,e^\prime,\varepsilon,\varepsilon^\prime}\mathbbm{f}_{m,\pi}^{ee^\prime\varepsilon\varepsilon^\prime}|ee^\prime\rangle\!\langle\varepsilon\varepsilon^\prime| \nonumber\\
    &= \prod_{n=1}^m(1-q_n) \mathbbm{f}_{m,\pi}^\markov \mbb{1}_\env + \sum_{\ell=0}^{m-1} \binom{q^{m-\ell}(1-q)^\ell\mathbbm{f}_{\ell,\pi}^\markov \hat{\mc{Q}}_{m-\ell,\pi}}{\hat{\mc{Q}}_{m-\ell,\pi}},
\end{align}
and thus the \gls{asf}~for this noise model with any finite group can be written as
\begin{align}
    \mc{F}_m &= \sum_{\pi\in{R}_\mathbb{G}}\llangle{M}|\hat{\tr}_\env\hat{\Gamma}_{m+1}(\hat{\mc{Q}}_{m,\pi}\otimes\hat{\mc{P}}_\pi)|\rho\rrangle \nonumber\\
    &= \prod_{n=1}^m(1-q_n)\!\sum_{\pi\in{R}_\mbb{G}}\mathbbm{f}_{m,\pi}^\markov \llangle{M}|\hat{\tr}_\env\hat{\Gamma}_{m+1}(\mbb1\otimes\hat{\mc{P}}_\pi)|\rho\rrangle \nonumber\\
    &\qquad\qquad+ \sum_{\ell=0}^{m-1} \sum_{\pi\in{R}_\mbb{G}}\llangle{M}|\hat{\tr}_\env\hat{\Gamma}_{m+1}\binom{q^{m-\ell}(1-q)^\ell\mathbbm{f}_{\ell,\pi}^\markov \hat{\mc{Q}}_{m-\ell,\pi}}{\hat{\mc{Q}}_{m-\ell,\pi}}\otimes\hat{\mc{P}}_\pi|\,\rho\rrangle,
\end{align}
where the first term is the single Markovian contribution, dominant whenever $q\approx0$.

For static noise, $\Gamma:=\Gamma_i = \Gamma_j$, this reduces to
\begin{align}
    &\mc{F}_m = \sum_{\pi\in{R}_\mathbb{G}}\llangle{M}|\hat{\tr}_\env\hat{\Gamma}(\hat{\mc{Q}}_{m,\pi}\otimes\hat{\mc{P}}_\pi)|\rho\rrangle \nonumber\\
    &= (1-q)^m\!\sum_{\pi\in{R}_\mbb{G}}f_\pi^m \llangle{M}|\hat{\tr}_\env\hat{\Gamma}(\mbb1\otimes\hat{\mc{P}}_\pi)|\rho\rrangle + \sum_{\ell=0}^{m-1}\binom{m}{\ell}\, q^{m-\ell}(1-q)^\ell\sum_{\pi\in{R}_\mbb{G}}f_\pi^\ell\llangle{M}|\hat{\tr}_\env\hat{\Gamma}(\hat{\mc{Q}}_{m-\ell,\pi}\otimes\hat{\mc{P}}_\pi)|\,\rho\rrangle \nonumber\\
    &= (1-q)^m\,\mc{F}_m^\markov + \sum_{\ell=0}^{m-1}\binom{m}{\ell} \,q^{m-\ell}(1-q)^\ell\,\mc{F}^{(\ell)}_{m},
    \label{eq: perturb nM}
\end{align}
which more easily can be read as a correction to the Markovian \gls{asf}~with contributions at each intermediate step. Here $\mc{F}_m^{(\ell)}$ means the \gls{asf}~decay is exponential (Markovian) up to timestep $\ell$, specifically given by
\begin{align}
    \mc{F}_{m}^{(\ell)} &:= \sum_{\pi\in{R}_{\mathbb{G}}}f_\pi^\ell \llangle{M}|\hat{\tr}_\env\hat{\Lambda}_{m+1}\left(\hat{\mc{Q}}_{m-\ell,\pi}\otimes\hat{\mc{P}}_\pi\right)|\rho\rrangle,
\end{align}
with $\hat{\mc{Q}}_{m-\ell,\pi}:=\sum{f}_{(\ell+1),\pi}^{ee^\prime\epsilon_{\ell+1}\epsilon^\prime_{\ell+1}}\cdots{f}_{(m),\pi}^{\epsilon_{m-1}\epsilon^\prime_{m-1}\varepsilon\varepsilon^\prime}|ee^\prime\rangle\!\langle\varepsilon\varepsilon^\prime|$ and $\rho$ being uncorrelated on $\syst\env$.

We can further relabel $\ell=m-\ell^\prime$ to obtain a binomial expansion dominated by the leading $\ell^\prime$ terms whenever $0\lesssim{q}\ll1/2$, i.e.
\begin{align}
    \mc{F}_m &= (1-q)^m\mc{F}_m^\markov + \sum_{\ell=1}^{m} \binom{m}{m-\ell}q^\ell(1-q)^{m-\ell}\mc{F}_m^{(m-\ell)}.
\end{align}
In this sense this constitutes a perturbative expansion with finite-memory corrections to the \gls{asf}.

\subsection{Character randomized benchmarking}\label{appendix: character rb}
The character of a group $\mbb{G}$ associated to the rep $\phi$ is a function $\chi_\phi:\mbb{G}\to\mbb{R}$. In particular, it satisfies
\begin{equation}
    \avg_{G\in\mbb{G}}\chi_\phi(G)\,\mc{G} = \f{1}{|\phi|}\hat{\mc{P}}_\phi,
\end{equation}
with $\hat{\mc{P}}_\phi$ the projector onto the support of all the subreps of $\mc{G}$ equivalent to $\phi$ and $|\phi|$ is the dimension of the representation $\phi$.

In particular, the core idea in the Character \gls{rb}~protocol of~\cite{Helsen2019} is that we can attach a gate from a subgroup of the \gls{rb}~group that we are benchmarking to isolate individual quality parameters.

Explicitly, fix a $\lambda^\prime\in{R}_{\mbb{G}}$; now consider $\mbb{H}\subset\mbb{G}$ such that the Liouville representation has a subrepresentation $\hat{\phi}$, with character function $\chi_{\hat{\phi}}$, that has support inside the rep $\phi_{\lambda^\prime}$ of $\mbb{G}$. This means $\hat{\mc{P}}_{\hat{\phi}}\subset\hat{\mc{P}}_{\lambda^\prime}$. Then attaching some extra gate $h\in\mbb{H}$ to the original \gls{rb}~sequence, we can estimate the weighed average
\begin{align}
    \kappa_m^{\lambda^\prime} &= |\hat{\phi}| \llangle{M}|\hat{\tr}_\env\hat{\Lambda}\left(\mbb{E}[(\mc{I}_\env\otimes\hat{\mc{G}}^\dg)\hat{\Lambda}(\mc{I}_\env\otimes\hat{\mc{G}})]\right)^m\!\avg_{h\in\mbb{H}}\chi_{\hat{\phi}}(h)\hat{h}|\rho\rrangle,\nonumber\\
    &= \sum_{e,e^\prime,\varepsilon,\varepsilon^\prime=1}^{\dimE}\sum_{\pi\in{R}_{\mathbb{G}}}\boldsymbol{\mathbbm{f}}_{m,\pi}^{ee^\prime\varepsilon\varepsilon^\prime}\llangle{M}\,|\hat{\tr}_\env\hat{\Lambda}\left(|ee^\prime\rangle\!\langle\varepsilon\varepsilon^\prime|\otimes\hat{\mc{P}}_\pi\hat{\mc{P}}_{\hat{\phi}}\right)|\,\rho\rrangle \nonumber\\
    &= \sum_{e,e^\prime,\varepsilon,\varepsilon^\prime=1}^{\dimE}\boldsymbol{\mathbbm{f}}_{m,\lambda^\prime}^{ee^\prime\varepsilon\varepsilon^\prime}\llangle{M}\,|\hat{\tr}_\env\hat{\Lambda}\left(|ee^\prime\rangle\!\langle\varepsilon\varepsilon^\prime|\otimes\hat{\mc{P}}_{\hat{\phi}}\right)|\,\rho\rrangle \nonumber\\
    &= \llangle{M}\,|\hat{\tr}_\env\hat{\Lambda}\left(\hat{\mc{Q}}_{m,\lambda^\prime}\otimes\hat{\mc{P}}_{\hat{\phi}}\right)|\,\rho\rrangle,
    \label{eq: character nM}
\end{align}
which isolates the part of the \gls{asf}~with quality tensor $\hat{\mc{Q}}_{m,\lambda^\prime}$. Whereas in the Markovian case one can readily fit single exponentials and extract average gate fidelities, for the non-Markovian case a further analysis or adapted protocol to extract the noise maps would be required, e.g., as we show, the averaging over initial states and measurements.

\section{Quality parameters and process fidelity}\label{appendix: process fidelity}
Single quality parameters arise from twirling via Schur's lemma, and the usual average gate fidelity of some channel $\Phi$ with the identity can also be directly related to a twirl. In fact, from the definition of the uniform average gate fidelity of a map $\Phi$ with respect to the identity,
\begin{align}
    \mathfrak{F}_{\Phi} &:= \int{d}\psi\langle\psi|\Phi(|\psi\rangle\!\langle\psi|)|\psi\rangle \nonumber\\
    &= \int{d}\mu(U)\, \langle\psi|U^\dg\Phi(U|\psi\rangle\!\langle\psi|{U}^\dg)U|\psi\rangle,
    \label{eq: second moment u}
\end{align}
where $\mu$ is the Haar measure and $U$ is a uniformly drawn random unitary, and given the formula
\begin{align}
    \int &d\mu(U)\,U^\dg A U X U^\dg B U = \f{d\tr(AB)-\tr(A)\tr(B)}{d(d^2-1)}\tr(X)\mbb1+\f{d\tr(A)\tr(B)-\tr(AB)}{d(d^2-1)}X,
    \label{eq: second moment}
\end{align}
for operators on a $d$-dimensional space, one can deduce that
\begin{equation}
    \mathfrak{F}_\Phi = \f{d+\tr[\hat{\Phi}]}{d(d+1)}.
    \label{eq: general avg gate fidelity}
\end{equation}

We may just employ invariance of the trace under a twirl with a subgroup $\mathbb{G}$ of the unitary group so that via Schur's lemma,
\begin{equation}
    \mathfrak{F}_\Phi = \f{d+\sum_{\pi\in{R}_{\mathbb{G}}}f_\pi\tr(\hat{\mc{P}}_\pi)}{d(d+1)}.
    \label{eq: Markov avg gate fidelity}
\end{equation}

In the standard case of Markovian \gls{rb}~with unitary 2-designs this gives a simple relation of the average gate fidelity with the noise-strength $p$ because the only quality parameters are $f_1=1$ and $f_2=p$, the latter of which contains the term $\tr[\Phi]$ such that
\begin{equation}
    \mathfrak{F}_\Phi = p + \f{1}{d}(1-p),
\end{equation}
and for this reason, in this scenario it suffices to run a standard \gls{rb}~protocol to extract average gate fidelities of the noise. More generally for finite groups (with a multiplicity-free representation) character \gls{rb}~described in Section~\ref{appendix: character rb} extracts each quality factor, allowing to estimate the respective average gate fidelity.

For the non-Markovian case, however, after an \gls{rb}~protocol, the quality tensor is still cloistered between the noise due to the undo gate and the state preparation error, both of which can act on the full $\syst\env$. While in the Markovian case these two can be absorbed as \gls{spam}~errors not affecting the \gls{asf}~decay, this is not the case for non-Markovian noise. Furthermore, non-Markovianity as a time-dependence is explicit in the quality map as all quality factors are correlated with each other through $\env$, as discussed in Appendix~\ref{appendix: time-dependent noise}.

Thus we propose quantifying the average fidelity of a noise process as a whole by means of \gls{rb}. Whenever this average fidelity factorizes, we will recover the standard Markovian case; this points to the possibility as well that other clear functional forms of this \emph{process fidelity} could be identified for certain types of noise. Consider then the following. Suppose we have a non-Markov \gls{asf}~given as in Eq.~\eqref{eq: non-Markov asf}; we care about how well the inputs are being mapped to outputs on average, thus let us define a map $\mathbf{F}_{m,\pi}^\syst$ such that
\begin{align}
    \mc{F}_m(\rho_\syst,M) = \sum_{\pi\in{R}_{\mbb{G}}}\tr\left[M\,\mathbf{F}_{m,\pi}^\syst(\rho_\syst)\right],
\end{align}
i.e., for some \gls{cptp}~map $\Lambda_0$ and a fiducial pure state $\varepsilon$ of $\env$, we have
\begin{align}
    \mathbf{F}_{m,\pi}^\syst(x) := \tr_\env\circ\Lambda_{m+1}\circ(\mc{Q}_{m,\pi}\otimes\mc{P}_\pi)\circ\Lambda_0(\varepsilon\otimes x).
\end{align}

The map $\Lambda_0$ thus explicitly refers to a state preparation error correlating the input solely in space $\syst$ with the environment $\env$. The previous discussion then means we wish to extract the average gate fidelity of the map $\mathbf{F}_{m,\pi}^\syst$. The \gls{asf}~contains only two free inputs which are chosen arbitrarily, $\rho_\syst$ and ${M}$: if we average over these, we could effectively extract the gate fidelity of $\mathbf{F}_{m,\pi}^\syst$. Of course, this added step in the \gls{rb}~protocol would only be relevant in the non-Markov case, i.e., when deviations from an exponential are observed, as otherwise it would simply amount of averaging the \gls{spam}~terms.

To extract an average gate fidelity we require a second moment over a given distribution, so consider
\begin{align}
    {M} &\to \mc{N}_{m+1}\circ\mc{U}(|s\rangle\!\langle{s}|) \\
    \rho_\syst &\to \mc{N}_0\circ\mc{U}(|\psi\rangle\!\langle\psi|),
\end{align}
where $\mc{U}$ is a unitary belonging to a 2-design, sampled at uniformly at random, $\{|s\rangle\}_{s=1}^{\dimS}$ is a basis vector of $\syst$, so that $M$ is the $s$\textsuperscript{th}-{\gls{povm}} element of such random {\gls{povm}}~\cite{heinosaari2019random}, $|\psi\rangle$ is a pure state in $\syst$, and each $\mc{N}_x$ is some \gls{cptp}~map modeling \gls{spam} due to the randomization. Overall we would now have a distribution of {\gls{asf}}s specified by
\begin{align}
    \mc{F}_m \to &\sum_{\pi\in{R}_{\mbb{G}}} \tr\left[\mc{N}_{m+1}\circ\mc{U}(|s\rangle\!\langle{s}|)\,\mathbf{F}_{m,\pi}^\syst\circ\mc{N}_0\circ\mc{U}(|\psi\rangle\!\langle\psi|)\right] \nonumber\\
    &= \sum_{\pi\in{R}_{\mbb{G}}} \langle{s}|\,\mc{U}^\dg\circ\mc{N}_{m+1}^\dg\circ\mathbf{F}_{m,\pi}^\syst\circ\mc{N}_0\circ\mc{U}(|\psi\rangle\!\langle\psi|)\,|s\rangle.
\end{align}

Then we can average over these inputs and measurements so that denoting by $K_\mu^{(m,\pi)}$ the Kraus operators of $\mathbf{F}_{m,\pi}^\syst$ and by $N^{(x)}_\mu$ the ones of $\mc{N}_x$, we get
\begin{align}
    \mbb{E}_{\rho_\syst,{M}}[\mc{F}_m] &= \sum \mbb{E}_U\langle{s}|\,U^\dg N^{\dg\,(m+1)}_\alpha K_\gamma^{(m,\pi)} N^{(0)}_\beta U|\psi\rangle\!\langle\psi|{U}^\dg N^{(0)\,\dg}_\beta K^{(m,\pi)\,\dg}_\gamma N^{(m+1)}_\alpha U\,|s\rangle \nonumber\\
    &:= \sum \mbb{E}_U\langle{s}|\,U^\dg \mathsf{K}_{\alpha\gamma\beta}^{(m,\pi)} U|\psi\rangle\!\langle\psi|{U}^\dg \mathsf{K}^{\dg\,(m,\pi)}_{\alpha\gamma\beta} U\,|s\rangle \nonumber\\
    &:= \sum_{\pi\in\mbb{G}} \mbb{E}_U\langle{s}|\,U^\dg \fid_{m,\pi}^\syst\left( U|\psi\rangle\!\langle\psi|{U}^\dg\right) U\,|s\rangle,
\end{align}
where we defined $\mathsf{K}_{\alpha\gamma\beta}^{(m,\pi)}:=N^{\dg\,(m+1)}_\alpha K_\gamma^{(m,\pi)} N^{(0)}_\beta$ as the Kraus operators of the composition
\begin{equation}
    \fid_{m,\pi}^{\syst} := \mc{N}_{m+1}^\dg\circ\mathbf{F}_{m,\pi}^\syst\circ\mc{N}_0.
\end{equation}

The maps $\fid_{m,\pi}^{\syst}$ and $\mathbf{F}_{m,\pi}^\syst$ are essentially equivalent, as we may simply absorb the $\mc{N}$ on the respective $\Lambda$ maps by defining
\begin{align}
    \boldsymbol{\mathsf{F}}_{m,\pi}^{\syst}(x) &= \tr_\env\circ(\mc{N}^\dg_{m+1}\!\!\circ\Lambda_{m+1})\circ(\mc{Q}_{m,\pi}\otimes\mc{P}_\pi)\circ(\Lambda_0\circ\mc{N}_0)(\varepsilon\otimes x)\nonumber\\
    &:= \tr_\env\circ\Lambda_{m+1}^\prime\circ(\mc{Q}_{m,\pi}\otimes\mc{P}_\pi)\circ\Lambda_0^\prime(\varepsilon\otimes x),
    \label{eq: def proc gate}
\end{align}
where the $\mc{N}$ maps act solely on $\syst$.

Notice that if $\langle{s}|\psi\rangle=1$, then $\mbb{E}_{\rho_\syst,M}[\mc{F}_m]=\sum_\pi\mathfrak{F}_{\fid}^{(m,\pi)}$ by definition, where here $\mathfrak{F}_{\fid}^{(m)}$ means the average gate fidelity of $\sum_\pi\fid_{m,\pi}^\syst$; we can now employ Eq.~\eqref{eq: second moment u} to get
\begin{align}
    \mbb{E}_{\rho_\syst,{M}}[\mc{F}_m] &= \sum_\pi\left[\f{\dimS\tr\left(\sum \mathsf{K}_{\alpha\gamma\beta}^{(m,\pi)}\mathsf{K}^{\dg\,(m,\pi)}_{\alpha\gamma\beta}\right) - \tr[\hat{\fid}_{m,\pi}^\syst]}{\dimS(\dimS^2-1)} + \f{\dimS\tr[\hat{\fid}_{m,\pi}^\syst] - \tr\left(\sum \mathsf{K}^{(m,\pi)}_{\alpha\gamma\beta}\mathsf{K}^{\dg\,(m,\pi)}_{\alpha\gamma\beta}\right)}{\dimS(\dimS^2-1)} |\langle{s}|\psi\rangle|^2\right].
\end{align}

 The full map $\fid_{m,\pi}^\syst$ sends $\syst$ input states to $\syst$ output states, so it must be {\gls{cp}}; we can now further demand the whole of $\sum_\pi\fid_{m,\pi}^\syst$ to be {\gls{cptp}}, so that $\tr[\sum_\pi\sum_{\alpha\beta\gamma}\mathsf{K}^{(m,\pi)}_{\alpha\gamma\beta}\mathsf{K}^{\dg\,(m,\pi)}_{\alpha\gamma\beta}]=\dimS$; this implies the noise being trace-preserving, then
 \begin{align}
    \mbb{E}_{\rho_\syst,{M}}[\mc{F}_m] &= \f{\dimS^2 - \sum_\pi\tr[\hat{\fid}_{m,\pi}^\syst]}{\dimS(\dimS^2-1)} + \f{\sum_\pi\tr[\hat{\fid}_{m,\pi}^\syst] - 1}{\dimS^2-1} |\langle{s}|\psi\rangle|^2 \nonumber\\
    &= \left(\f{\sum_\pi\tr[\hat{\fid}_{m,\pi}^\syst] - 1}{\dimS^2-1}\right)\left(|\langle{s}|\psi\rangle|^2-\f{1}{\dimS}\right)+\left(\f{1}{\dimS}\right) \nonumber\\
    &= \left(\f{\dimS\,\mathfrak{F}^{(m)}_{\fid}-1}{\dimS-1}\right)\left(|\langle{s}|\psi\rangle|^2-\f{1}{\dimS}\right)+\left(\f{1}{\dimS}\right)\nonumber\\
    &= \left(\f{\dimS|\langle{s}|\psi\rangle|^2-1}{\dimS-1}\right) \mathfrak{F}^{(m)}_{\fid} + \f{1-|\langle{s}|\psi\rangle|^2}{\dimS-1}\nonumber\\
    &:= A\mathfrak{F}^{(m)}_{\fid} + B
    \label{eq: avg states and M}
\end{align}
where the second line can be identified noticing that $\f{\dimS^2 - \sum_\pi\tr[\hat{\fid}_{m,\pi}^\syst]}{\dimS^2-1} = 1-\f{\sum_\pi\tr[\hat{\fid}_{m,\pi}^\syst] - 1}{\dimS^2-1}$, the third line follows by Eq.~\eqref{eq: general avg gate fidelity}, and $A$ and $B$ are constants. The form of the average in Eq.~\eqref{eq: avg states and M} follows from it being uniform with respect to either the full unitary group or a 2-design, while the gate fidelity $\mathfrak{F}_{\fid}^{(m)}$ is still a quantity containing the \gls{rb}~protocol, i.e., average noise, with respect to any finite group.

The dependence in $m$ will generally be non-trivial, as it is directly related to a dependence of the environment $\env$, except for specific cases or whenever there are assumptions on the noise, e.g., for Markov time-independent noise, and other particular cases might also display some useful functional structure.

Consider the definition in Eq.~\eqref{eq: def proc gate} and let $\hat{\mathfrak{S}}_{\varepsilon}:=(|\varepsilon\rangle\otimes\mbb1_\syst)\otimes(|\varepsilon\rangle\otimes\mbb1_\syst)^*$, then we have
\begin{align}
    \sum_\pi\tr[\hat{\fid}_{m,\pi}^\syst] &= \sum_\pi\tr\left[\hat{\tr}_\env\hat{\Lambda}^\prime_{m+1}(\hat{\mc{Q}}_{m,\pi}\otimes\hat{\mc{P}}_\pi)\hat{\Lambda}^\prime_0\hat{\mathfrak{S}}_{\varepsilon}\right] \nonumber\\
    &= \tr\left[\hat{\Lambda}^\prime_0\hat{\mathfrak{S}}_{\varepsilon}\hat{\tr}_\env\hat{\Lambda}^\prime_{m+1}\,\sum_\pi(\hat{\mc{Q}}_{m,\pi}\otimes\hat{\mc{P}}_\pi)\right] \nonumber\\
    &\leq \|\hat{\Lambda}^\prime_0\hat{\mathfrak{S}}_{\varepsilon}\hat{\tr}_\env\hat{\Lambda}^\prime_{m+1}\| \sum_\pi\tr\left(\hat{\mc{Q}}_{m,\pi}\right)\tr\left(\hat{\mc{P}}_\pi\right) \nonumber\\
    &:= \|\hat{\mscr{S}}\|\sum_\pi\tr\left(\hat{\mc{Q}}_{m,\pi}\right)\tr\left(\hat{\mc{P}}_\pi\right),
\end{align}
where here $\|\cdot\|$ denotes operator norm, i.e., largest singular value, and for the first inequality we used the fact that all operators are positive semidefinite. Finally, we defined $\hat{\mscr{S}}$ as the \gls{spam}-noise operator $\hat{\mscr{S}}:=\hat{\Lambda}^\prime_0\hat{\mathfrak{S}}_{\varepsilon}\hat{\tr}_\env\hat{\Lambda}^\prime_{m+1}$; in terms of Kraus operators, this is
\begin{align}
    \hat{\mscr{S}} = \sum_{\mu,\nu}\sum_e \left(\lambda_0^\mu|\varepsilon\rangle\!\langle{e}|\lambda_{m+1}^\nu\otimes\lambda_0^{*\mu}|\varepsilon\rangle\!\langle{e}|^*\lambda_{m+1}^{*\nu}\right).
\end{align}

Now we have
\begin{align}
    \sum_\pi\tr\left(\hat{\mc{Q}}_{m,\pi}\right)\tr\left(\hat{\mc{P}}_\pi\right) &\geq \f{\dimS[(\dimS+1)\mathfrak{F}_{\fid}^{(m)} - 1]}{\|\hat{\mscr{S}}\|}  ,
\end{align}
so that Character \gls{rb} can be employed to obtain estimates of $\tr\left(\hat{\mc{Q}}_{m,\pi}\right)\tr\left(\hat{\mc{P}}_\pi\right)$, which is related to the gate fidelity of the map $\mc{Q}_{m,\pi}\otimes\mc{P}_\pi$, as in Eq.~\eqref{eq: general avg gate fidelity}. This can serve to estimate noise from gates separately from \gls{spam} correlation errors.

\subsection{Process fidelity - The Markovian case}\label{appendix: avg process fidelity Markov}
Clearly when averaging over initial states and measurements, the \gls{asf} of a Markovian \gls{rb} experiment should simply average the \gls{spam} terms. This might still be useful in case average \gls{spam} error rates are of interest. First, Eq.~\eqref{eq: def proc gate} turns into
\begin{align}
    \boldsymbol{\mathsf{F}}_{m,\pi}^{\syst}
    &:= \left(\prod_{i=1}^mf_{(i),\pi}\right)\,\Lambda_{m+1}^\prime\circ\mc{P}_\pi\circ\Lambda_0^\prime,
\end{align}
with all maps now acting solely on $\syst$. Then,
 \begin{align}
    \mbb{E}_{\rho_\syst,{M}}[\mc{F}_m] &= 
    \sum_{\pi\in{R}_{\mbb{G}}}\left(\prod_{i=1}^mf_{(i),\pi}\right) \mbb{E}_U \tr\left[U|s\rangle\!\langle{s}|U^\dg\Lambda_{m+1}^\prime\circ\mc{P}_\pi\circ\Lambda_0^\prime(U|\psi\rangle\!\langle\psi|U^\dg)\right]\nonumber\\
    &= \sum_{\pi\in{R}_{\mbb{G}}}\left(\prod_{i=1}^mf_{(i),\pi}\right) \mbb{E}_{M,\rho_{_\syst}} \tr\left[M_s\Lambda_{m+1}\circ\mc{P}_\pi\circ\Lambda_0(\rho_{_\syst})\right]\nonumber\\
    &= \sum_{\pi\in{R}_{\mbb{G}}}\left(\prod_{i=1}^mf_{(i),\pi}\right) \mbb{E}_{M,\rho_{_\syst}} \llangle M_s|\hat{\Lambda}_{m+1}\hat{\mc{P}}_\pi\hat{\Lambda}_0|\rho_{_\syst}\rrangle,
\end{align}
which amounts exactly to uniformly averaging the \gls{spam}~terms over noisy measurement elements and initial states.

Now, however, we have
\begin{align}
    \mathfrak{F}_{\fid}^{(m)} &= \f{\dimS + \sum_\pi\tr\left(\hat{\fid}_{m,\pi}^\syst\right)}{\dimS(\dimS+1)} \nonumber\\
    &= \f{\dimS + \sum_\pi\left(\prod_{i=1}^mf_{(i),\pi}\right)\tr\left(\hat{\Lambda}_0^\prime\hat{\Lambda}_{m+1}^\prime\hat{\mc{P}}_\pi\right)}{\dimS(\dimS+1)}, \nonumber\\
    &= \f{\dimS + \sum_\pi\left(\prod_{i=1}^mf_{(i),\pi}\right)f_\pi^{(\Lambda_0^\prime\Lambda_{m+1}^\prime)}\tr\left(\hat{\mc{P}}_\pi\right)}{\dimS(\dimS+1)}
\end{align}
where $f_\pi^{(\Lambda_0^\prime\Lambda_{m+1}^\prime)}$ is a quality factor of the \gls{spam} noise. Here clearly all of the $m$-steps are being taken into account and the \gls{spam} errors are directly incorporated: this is because the average over initial states and measurements will always contain both the $\Lambda_0$ and $\Lambda_{m+1}$ maps.

For the unitary 2-design / Clifford case, with time-independence and \gls{cptp} noise, it can be seen that
\begin{align}
    \mbb{E}_{\rho_\syst,{M}}[\mc{F}_m] &= p^m q\left(|\langle{s}|\psi\rangle|^2-\f{1}{\dimS}\right)+\f{1}{\dimS},
\end{align}
where
\begin{align}
    q &= \f{\tr[\hat{\Lambda}_0^\prime\hat{\Lambda}_{m+1}^\prime]-1}{\dimS^2-1},
\end{align}
being the corresponding \gls{spam} quality factor or noise-strength.

\section{Non-Markovianity}\label{appendix: non-Markov}
Now, looking back at quality factors, we have
\begin{align}
    \tr[\hat{\mc{Q}}_{m,\pi}] &= \sum_{\{e,\epsilon\}}{f}_{(1),\pi}^{ee^\prime\epsilon_1\epsilon_1^\prime}{f}_{(2),\pi}^{\epsilon_1\epsilon_1^\prime\epsilon_2\epsilon_2^\prime}{f}_{(3),\pi}^{\epsilon_2\epsilon_2^\prime\epsilon_3\epsilon_3^\prime}\cdots{f}_{(m),\pi}^{\epsilon_{m-1}\epsilon_{m-1}^\prime ee^\prime}.
\end{align}

In the Markovian time-dependent case, it is possible to estimate average gate (in)fidelities on arbitrary sequence length intervals~\cite{Wallman_2014} by taking ratios of {\gls{asf}}s, precisely because these do not depend on all previous steps, i.e., the quality factors satisfy $\f{f_{(1)}f_{(2)}\cdots{f}_{(m_2)}}{f_{(1)}f_{(2)}\cdots{f}_{(m_1)}}=f_{(m_1+1)}\cdots{f}_{(m_2)}$ for $m_2>m_1$. As we argue in Appendix~\ref{appendix: time-dependent noise}, time-dependent Markovian {\gls{asf}}s are contained in non-Markovian ones, either time-independent or time-dependent, a crucial difference, however, is that quality factors in time-dependent Markovian {\gls{asf}}s only change its rate of decay, so it is not possible to obtain a higher {\gls{asf}} in increasing sequence lengths, while there is virtually no reason why non-Markovian {\gls{asf}}s should also be constrained in this way.

Consider then $n>m$ such that $\mc{F}_n>\mc{F}_m$ for \gls{rb} experiments with the exact same group and noise up to step $m$. This would imply that
\begin{align}
    \tr\left\{|\rho_\env\otimes\rho_\syst\rrangle\!\!\llangle{M}|\hat{\tr}_\env\sum_\pi\left[\hat{\Lambda}_{n+1}\left(\hat{\mc{Q}}_{n,\pi}\otimes\mbb1_\syst\right)-\hat{\Lambda}_{m+1}\left(\hat{\mc{Q}}_{m,\pi}\otimes\mbb1_\syst\right)\right]\left(\mbb1_\env\otimes\hat{\mc{P}}_\pi\right)\hat{\Lambda}_0\right\}>0,
\end{align}
which will hold whenever
\begin{align}
    \hat{\Lambda}_{n+1}\left(\hat{\mc{Q}}_{n,\pi}\otimes\mbb1\right)-\hat{\Lambda}_{m+1}\left(\hat{\mc{Q}}_{m,\pi}\otimes\mbb1\right) \succ 0
\end{align}
i.e. where the matrix difference on the left side is positive definite for all $\pi$, in turn implying
\begin{align}
    \tr\left[\tr_\syst\left(\hat{\Lambda}_{n+1}\right)\hat{\mc{Q}}_{n,\pi}\right]-\tr\left[\tr_\syst\left(\hat{\Lambda}_{m+1}\right)\hat{\mc{Q}}_{m,\pi}\right]>0,
\end{align}
or equivalently, letting $\hat{\Lambda}_i^\env:=\tr_\syst(\hat{\Lambda}_i)$,
\begin{align}
    \sum_\env\,\mathbbm{f}_{m-1,\pi}^{ee^\prime\epsilon_{m-1}\epsilon_{m-1}^\prime}\left(\mathbbm{f}_{n-m+1, \pi}^{\epsilon_{m-1}\epsilon_{m-1}^\prime \varepsilon\varepsilon^\prime}\langle\varepsilon\varepsilon^\prime|\hat{\Lambda}_{n+1}^\env|ee^\prime\rangle - f_{(m),\pi}^{\epsilon_{m-1}\epsilon_{m-1}^\prime \varepsilon\varepsilon^\prime}\langle\varepsilon\varepsilon^\prime|\hat{\Lambda}_{m+1}^\env|ee^\prime\rangle\right) &> 0,
\end{align}
where the sum here is over all environment $e,\epsilon$ indices. Let us first consider the simplest case, $n = m+1$, so that
\begin{align}
    \sum_\env\,\mathbbm{f}_{m-1,\pi}^{ee^\prime\epsilon_{m-1}\epsilon_{m-1}^\prime}\left(f_{(m),\pi}^{\epsilon_{m-1}\epsilon_{m-1}^\prime\epsilon_m\epsilon_m^\prime}f_{(m+1),\pi}^{\epsilon_{m}\epsilon_{m}^\prime\varepsilon\varepsilon^\prime}\langle\varepsilon\varepsilon^\prime|\hat{\Lambda}_{m+2}^\env|ee^\prime\rangle - f_{(m),\pi}^{\epsilon_{m-1}\epsilon_{m-1}^\prime \varepsilon\varepsilon^\prime}\langle\varepsilon\varepsilon^\prime|\hat{\Lambda}_{m+1}^\env|ee^\prime\rangle\right) &> 0,
\end{align}
and furthermore, consider finite memory such that effectively $m=1$, i.e.
\begin{align}
    \sum_\env\left(f_{(1),\pi}^{ee^\prime\epsilon_1\epsilon_1^\prime}f_{(2),\pi}^{\epsilon_{1}\epsilon_{1}^\prime \varepsilon\varepsilon^\prime}\langle\varepsilon\varepsilon^\prime|\hat{\Lambda}_3^\env|ee^\prime\rangle - f_{(1),\pi}^{ee^\prime\varepsilon\varepsilon^\prime}\langle\varepsilon\varepsilon^\prime|\hat{\Lambda}_{2}^\env|ee^\prime\rangle\right) &> 0,
\end{align}
which by definition means
\begin{align}
    0 <& \sum_\env\left(\f{\tr\left[\langle{ee^\prime\epsilon_1\epsilon_1^\prime}|(\hat{\Lambda}_1\otimes\hat{\Lambda}_2)\hat{\Lambda}_3^\env|\epsilon_1\epsilon_1^\prime ee^\prime\rangle\hat{\mc{P}}_\pi^{\otimes2}\right]}{\tr\left(\hat{\mc{P}}_\pi^{\otimes2}\right)} - \f{\tr\left[\langle{ee^\prime}|(\hat{\Lambda}_1)\hat{\Lambda}_2^\env|ee^\prime\rangle\hat{\mc{P}}_\pi\right]}{\tr\left(\hat{\mc{P}}_\pi\right)}\right) \nonumber\\
    &= \tr\left(\f{\hat{\Lambda}_1(\mbb1_\env\otimes\hat{\mc{P}}_\pi)\hat{\Lambda}_2(\hat{\Lambda}_3^\env\otimes\hat{\mc{P}}_\pi)}{\tr\left(\hat{\mc{P}_\pi}^{\otimes2}\right)} - \f{\hat{\Lambda}_1(\hat{\Lambda}_2^\env\otimes\hat{\mc{P}}_\pi)}{\tr\left(\hat{\mc{P}_\pi}\right)}\right) \nonumber\\
    &= \f{1}{\tr\left(\hat{\mc{P}}_\pi^{\otimes2}\right)} \tr\left\{\hat{\Lambda}_1\left[(\mbb1_\env\otimes\hat{\mc{P}}_\pi)\hat{\Lambda}_2(\hat{\Lambda}_3^\env\otimes\hat{\mc{P}}_\pi) - (\hat{\Lambda}_2^\env\otimes\hat{\mc{P}}_\pi)\tr\left(\hat{\mc{P}}_\pi\right)\right] \right\} \nonumber\\
    &\leq \f{\|\hat{\Lambda}_1(\mbb1_\env\otimes\hat{\mc{P}}_\pi)\|}{\tr\left(\hat{\mc{P}}_\pi^{\otimes2}\right)} \tr\left[\hat{\Lambda}_2(\hat{\Lambda}_3^\env\otimes\hat{\mc{P}}_\pi) - (\hat{\Lambda}_2^\env\otimes\mbb1_\syst)\tr\left(\hat{\mc{P}_\pi}\right)\right],
    \label{ineq: nM quality Lambda}
\end{align}
and so,
\begin{align}
    \dimS\tr(\hat{\Lambda}_2)\tr\left(\hat{\mc{P}}_\pi\right) < \tr\left[\hat{\Lambda}_2\left(\hat{\Lambda}_3^\env\otimes\hat{\mc{P}}_\pi\right)\right],
\end{align}
where furthermore, we may use $\tr(XY)\leq\|X\|\tr(Y)$ for positive semidefinite $X$, $Y$, on the right-hand-side, so that
\begin{align}
    \tr(\hat{\Lambda}_3)\|\hat{\Lambda}_2\|>\dimS\tr(\hat{\Lambda}_2).
\end{align}

The general form of inequality~\eqref{ineq: nM quality Lambda} is now clear for any $m$ and $n$ as
\begin{align}
    \tr\left[\hat{\Lambda}_{m+1}(\mbb1\otimes\hat{\mc{P}}_\pi)\cdots\hat{\Lambda}_{n}(\hat{\Lambda}_{n+1}^\env\otimes\hat{\mc{P}}_\pi)-(\hat{\Lambda}_{m+1}^\env\otimes\mbb1)\tr(\hat{\mc{P}}_\pi)\right] > 0,
\end{align}
simplified to
\begin{equation}
     \|\hat{\Lambda}_{m+1}\|\cdots\|\hat{\Lambda}_n\|\tr(\hat{\Lambda}_{n+1}) > \dimS\tr(\hat{\Lambda}_{m+1}),
\end{equation}
repeatedly employing $\tr(XY)\leq\|X\|\tr(Y)$ for positive semidefinite $X$, $Y$. Both trace terms are lower-bounded by zero, although they realistically are less and close to $\dimE\dimS$.

In particular, for time-independent noise this reduces to
\begin{align}
    \|\hat{\Lambda}\|^{n-m} > \dimS,
\end{align}
which says that $\Lambda$ must be able to increase the purity of its input in at least $\dimS^{2/(n-m)}$, ruling out e.g., coherent noise. Furthermore, by $\|\hat{\mc{X}}\|\leq\sqrt{d}$ for \gls{cptp} map $\mc{X}$ acting between $d$-dimensional spaces (theorem II.I in~\cite{OpNormCPTP_2006}), one gets $\dimE>\dimS^{\f{2}{n-m}-1}$, which really just implies the necessity of an environment.


\section{Gate-dependence}\label{appendix: gate-dependent noise}
\subsection{General case --- gate-dependence is taken to the environment}\label{appendix: gate-dep wallman}
Suppose now instead of having chosen to model the noisy gates through the maps $\Lambda$, we choose to employ $\mc{J}:= \mc{L}\circ\mc{G}\circ\mc{R}$ for some given {\gls{cp}} maps $\mc{L}$ and $\mc{R}$. This is such that a time-independent and gate-independent \gls{rb} sequence reads
\begin{align}
    &\mc{S}_m = \Mcirc_{i=1}^{m+1}\mc{J}_i \nonumber\\
    &=\mc{L}\circ\mc{G}_{m+1}\circ(\mc{R}\circ\mc{L})\circ\mc{G}_m\circ(\mc{R}\circ\mc{L})\circ\cdots\circ\mc{G}_1\circ\mc{R},
\end{align}
which produces an equivalent \gls{asf} as the one using the noise maps $\Lambda$, here with the outermost noisy maps accounting for \gls{spam} errors. We may now incorporate gate-dependence by taking
\begin{align}
    \mc{J}^{(g)} &:= \mc{J} + \Delta_g \nonumber\\
    &= \mc{L}\circ\mc{G}\circ\mc{R} + \Delta_g,
    \label{eq: gate dep noise}
\end{align}
where $\Delta_g$ is the map containing the gate-dependent contribution, which is not necessarily small. In the non-Markovian case, $\mc{L}$, $\mc{R}$, $\Delta_g$, and $\mc{J}$ all act on both $\syst\env$, while the ideal $\mc{G}$ acts on $\syst$ alone. 

Let us now denote $\mc{X}_{j:i} = \mc{X}_j\circ\cdots\circ\mc{X}_i$, and $X_{j:i} = X_j\cdots{X}_i$ for any maps $\mc{X}_i,\ldots,\mc{X}_j$ or matrices $X_i,\ldots,X_j$. We can now expand the corresponding noisy sequence as
\begin{align}
    \mc{S}_m^{(g)} &= \mc{J}^{(g)}_{m+1:1} \nonumber\\
    &= \sum_{\{\ell_i=0\}}^1 \mc{J}_{m+1}^{\ell_{m+1}}\circ\Delta_{m+1}^{1-\ell_{m+1}}\circ\mc{J}_{m}^{\ell_{m}}\circ\Delta_{m}^{1-\ell_{m}}\circ\cdots\circ\mc{J}_{1}^{\ell_1}\circ\Delta_{1}^{1-\ell_1} \nonumber\\
    &= \mc{J}_{m+1:1} + \Delta_{m+1:1} + \kern-2em \sum_{\substack{\{\ell_i=0\}\\\setminus\{\text{all}\,\ell_i=0\,\text{or}\,\text{all}\,\ell_i=1\}}}^1 \kern-2em \mc{J}_{m+1}^{\ell_{m+1}}\circ\Delta_{m+1}^{1-\ell_{m+1}}\circ\mc{J}_{m}^{\ell_{m}}\circ\Delta_{m}^{1-\ell_{m}}\circ\cdots\circ\mc{J}_{1}^{\ell_1}\circ\Delta_{1}^{1-\ell_1},
    \label{eq: gate-dep sequence}
\end{align}
where here $\Delta_i=\Delta_{g_i}$, where the second line follows from the \href{https://en.wikipedia.org/wiki/Binomial_theorem#Multi-binomial_theorem}{multi-binomial theorem} and the last line contains terms mixing $\mc{J}$ and $\Delta$: the results in~\cite{Wallman_2018, Helsen2019} imply that for Markovian noise, such mixed terms do not contribute to the \gls{asf}, and that the term $\Delta_{m+1:1}$ gives a contribution that vanishes exponentially in sequence length.

Extending the previous result to a non-Markovian \gls{asf} in gate-dependent noise would require $\mc{L}$ and $\mc{R}$ to satisfy the properties
\begin{align}
    \mbb{E}\left[\hat{\mc{J}}^{(g)}\hat{\mc{L}}\,\hat{\mc{G}}^\dg\right] &= \hat{\mc{L}}\,\hat{\mc{D}}_g,
    \label{eq: app gate dep left} \\
    \mbb{E}\left[\hat{\mc{G}}^\dg\hat{\mc{R}}\,\hat{\mc{J}}^{(g)}\right] &= \hat{\mc{D}}_g \hat{\mc{R}},
    \label{eq: app gate dep right}\\
    \mbb{E}\left[\hat{\mc{G}}\,\hat{\mc{R}}\hat{\mc{L}}\,\hat{\mc{G}}^\dg\right] &= \hat{\mc{D}}_g,
    \label{eq: app gate dep mid}
\end{align}
where averages again are uniform over the group $\mathbb{G}$, with
\begin{equation}
    \hat{\mc{D}}_g := \sum_{\pi\in{R}_\mathbb{G}} \hat{\mscr{Q}}_\pi\otimes \hat{\mc{P}}_\pi,
\end{equation}
for some $\env$ to $\env$ maps $\mscr{Q}_\pi$.

Following~\cite{Helsen2019}, start by plugging the definition of $\mc{D}_g$ in Eq.~\eqref{eq: app gate dep left} and Eq.~\eqref{eq: app gate dep right}, together with the multiplicity-free decomposition of the Liouville representation $\hat{\mc{G}}:=\displaystyle{\Moplus_{\pi\in{R}_{\mathbb{G}}}\phi_\pi(g)}$, so that
\begin{align}
    \sum_\pi\mbb{E}\left[\hat{\mc{J}}^{(g)}\,\hat{\mc{L}}\,\hat{\mc{P}}_\pi\,\phi_\pi(g)^\dg\right] &= \sum_\pi\hat{\mc{L}}\,(\hat{\mscr{Q}}_\pi\otimes \hat{\mc{P}}_\pi), \\
    \sum_\pi\mbb{E}\left[\phi_\pi(g)^\dg\,\hat{\mc{P}}_\pi\,\hat{\mc{R}}\,\hat{\mc{J}}^{(g)}\right] &= \sum_\pi (\hat{\mscr{Q}}_\pi\otimes \hat{\mc{P}}_\pi)\,\hat{\mc{R}}.
\end{align}

Now the difference is that both $\hat{\mc{P}}_\pi$ and $\phi(g)$ act solely on $\syst$, i.e., we now take w.l.o.g.,
\begin{align}
    \hat{\mc{L}}:=\sum_\pi\hat{\mc{L}}_\pi,\qquad \hat{\mc{L}}_\pi\left(\mbb1_\env\otimes\hat{\mc{P}}_\lambda\right) = \delta_{\pi\lambda}\hat{\mc{L}}_\pi,\,\forall\lambda, \label{eq: app L decomp nM}\\
    \hat{\mc{R}}:=\sum_\pi\hat{\mc{R}}_\pi,\qquad \left(\mbb1_\env\otimes\hat{\mc{P}}_\lambda\right)\,\hat{\mc{R}}_\pi = \delta_{\pi\lambda}\hat{\mc{R}}_\pi,\,\forall\lambda,\label{eq: app R decomn nM}
\end{align}

That is, writing the identities explicitly, then we have the equations
\begin{align}
    \mbb{E}\left[\hat{\mc{J}}^{(g)}\,\hat{\mc{L}}_\pi\,(\mbb1_\env\otimes\phi_\pi(g)^\dg)\right] &= \hat{\mc{L}}_\pi\,(\hat{\mscr{Q}}_\pi\otimes\mbb1_\syst), \\
    \mbb{E}\left[(\mbb1_\env\otimes\phi_\pi(g)^\dg)\,\hat{\mc{R}}_\pi\,\hat{\mc{J}}^{(g)}\right] &= (\hat{\mscr{Q}}_\pi\otimes\mbb1_\syst)\,\hat{\mc{R}}_\pi,
\end{align}
and we can vectorize both sides to get
\begin{align}
    \mbb{E}\left[\hat{\mc{J}}^{(g)}\otimes\mbb1_\env\otimes\phi_\pi(g)^*\right]|\,\hat{\mc{L}}_\pi\rrangle&= (\mbb1_{\env\syst}\otimes\hat{\mscr{Q}}_\pi^{\,\mathrm{T}}\otimes\mbb1_\syst)\,|\hat{\mc{L}}_\pi\rrangle, \\
    \mbb{E}\left[(\mbb1_\env\otimes\phi_\pi(g)^*\otimes\hat{\mc{J}}^{(g)})^\mathrm{T}\right] |\hat{\mc{R}}_\pi\rrangle &= (\hat{\mscr{Q}}_\pi\otimes\mbb1_\syst\otimes\mbb1_{\env\syst})|\hat{\mc{R}}_\pi\rrangle,
\end{align}
or equivalently (transposing and reordering spaces),
\begin{align}
    \llangle\hat{\mc{L}}_\pi|\left\{\mbb1_\env\otimes\mbb{E}\left[\left(\phi_\pi(g)^*\otimes\hat{\mc{J}}^{(g)}\right)^{\mathrm{T}}\right]\right\} &= \llangle\hat{\mc{L}}_\pi|(\hat{\mscr{Q}}_\pi\otimes\mbb1_{\syst\env\syst}), \\
    \left\{\mbb1_\env\otimes\mbb{E}\left[\left(\phi_\pi(g)^*\otimes\hat{\mc{J}}^{(g)}\right)^{\mathrm{T}}\right]\right\} |\hat{\mc{R}}_\pi\rrangle &= (\hat{\mscr{Q}}_\pi\otimes\mbb1_{\syst\env\syst})|\hat{\mc{R}}_\pi\rrangle.
\end{align}

In the Markov case, these are simply eigenvalue equations (as the quality factor is a scalar), so the solution of $\mc{L}$ and $\mc{R}$ follows through them being eigenvectors of the average operator in the left-hand side.

\subsection{Stability under gate-dependent perturbations}\label{appendix: gate-dep stability}
While this is not possible in the non-Markovian case, we may consider a perturbative gate-dependence and look at how the \gls{asf} and the variance of the sequence fidelity changes. As we now show, this can be done in a similar way as done for the Markov and Clifford case in~\cite{Wallman_2014}. That is, take the noisy gates to be
\begin{align}
    \mc{J}^{(g)} = \mc{J}+ \epsilon \Delta_g = \Lambda\circ\mc{G} + \epsilon \Delta_g,
\end{align}
for some $\epsilon$ scaled such that $\|\hat{\Delta}_g\|\leq1$ for all $g$, and denote the sequence fidelity by $\mc{Z}_m^{(g)}:=\llangle M|\hat{\mc{J}}^{(g)}_{m+1:1}|\rho\rrangle$ for fixed sequence length $m$. Due to Eq.~\eqref{eq: gate-dep sequence}, we have
\begin{align}
    \mc{Z}_m^{(g)} =  \mc{Z}_m + \sum_{n=1}^{m+1} \epsilon^n \, \llangle{M}|\,\hat{\Xi}_m^{(n)}|\,\rho\rrangle,
    \label{eq: seq fid gate-preturbation}
\end{align}
where $\mc{Z}_m$ is the gate-independent sequence fidelity and where here $\Xi_m^{(n)}$ contains gate-dependent contributions to the $n$\textsuperscript{th} order in $\epsilon$, i.e., all terms containing $n$ amount of $\Delta$ terms in the gate-sequence; specifically we can write
\begin{align}
    \Xi_m^{(n)} := \sum_{s\in\mathbb{Z}_2^{m+1}:H(s)=n}\prod_{i=m+1}^{1}\mc{W}^{(g)}_{i,s},\quad\text{where}\quad\mc{W}^{(g)}_{i,s}=\begin{cases}\mc{J}_i,&\text{if}\,s=0 \\ \Delta_i,&\text{if}\,s=1\end{cases}
\end{align}
which is equivalent to the second and third terms of Eq.~\eqref{eq: gate-dep sequence}, just highlighting the amount $n$ of $\Delta$ terms; here $H(s)$ is the number of bits equal to 1 in the bit string $s$ (more generally, its Hamming weight) and $\Delta_i:=\Delta_{g_i}$. Now then, the gate-dependent \gls{asf}, $\mc{F}_m^{(g)}$, has the form
\begin{align}
    \mc{F}_m^{(g)} = \mc{F}_m + \sum_{n=1}^{m+1} \epsilon^n \, \llangle{M}|\mbb{E}\left[\hat{\Xi}_m^{(n)}\right]|\,\rho\rrangle,
\end{align}
and the gate-dependent term satisfies the following,
\begin{align}
    \left|\sum_{n=1}^{m+1}\epsilon^n \llangle{M}|\mbb{E}\left[\hat{\Xi}_m^{(n)}\right]|\,\rho\rrangle\right| &\leq \sum_{n=1}^{m+1}\epsilon^n\, \left\|\mbb{E}\left[\hat{\Xi}_m^{(n)}\right]\right\| \nonumber\\
    &\leq \sum_{n=1}^{m+1}\epsilon^n\, \sum_{s:H(s)=n}\|\mbb{E}\left[\prod_{i=m+1}^1\mc{W}^{(g)}_{i,s}\right]\|\nonumber\\
    &\leq \sum_{n=1}^{m+1}\epsilon^n\,\binom{m+1}{n}\,(\dimE\dimS)^{(n+1)/2} \nonumber\\
    &= \sqrt{\dimE\dimS}\sum_{n=1}^{m+1}\,\binom{m+1}{n}\,\left(\epsilon\sqrt{\dimE\dimS}\right)^{n} \nonumber\\
    &= \sqrt{\dimE\dimS} \left[\left(1 + \epsilon\sqrt{\dimE\dimS}\right)^{m+1} - 1\right] \nonumber\\
    &\leq \sqrt{\dimE\dimS}\left[\ex^{\epsilon\sqrt{\dimE\dimS}(m+1)}-1\right],
    \label{eq: bound perturb}
\end{align}
where $\|\cdot\|$ denotes maximum singular value (operator norm) of the Liouville representation of $\Xi_m^{(n)}$, and on the third line we use $\|\hat{\mc{X}}\|\leq\sqrt{d}$ for a \gls{cptp}~map between $d$-dimensional spaces (theorem II.I in~\cite{OpNormCPTP_2006}), together with $\|AB\|\leq\|A\|\,\|B\|$, which using the assumption that $\|\Delta_i\|\leq1$, gives only $(n+1)$ terms $\|\hat{\mc{J}}\|$. We point out there is virtually no change from the derivation of~\cite{Wallman_2014}: in particular, the average in the second line and/or the fact that the gates act solely on $\syst$, do not change anything, and the bound is (almost) the same derived there with $d\to\dimE\dimS$; this rather has to do with the noise acting on the full $\syst\env$, rather than \gls{rb} being done solely on $\syst$.

From here, letting $\delta_{\mc{F}}:=|\mc{F}_m^{(g)}-\mc{F}_m|$ we get
\begin{align}
    \epsilon \leq \f{\log\left[1 + \delta_{\mc{F}}/\sqrt{\dimE\dimS}\right]}{(m+1)\sqrt{\dimE\dimS}},
\end{align}
that is, we can say that $|\mc{F}_m^{(g)}-\mc{F}_m|\leq\delta$ whenever $\epsilon\leq\log\left[1 + \delta\,(\dimE\dimS)^{-1/2}\right](m+1)^{-1}(\dimE\dimS)^{-1/2}$. In the main text we further use $\log(1+x) \leq x$ for $x>-1$, or in particular $\log(1+x) \approx x$ for small $x$.

Turning now to the variance $\mbb{V}$ of the sequence fidelity, $\mc{Z}_m^{(g)}$, the way to obtain the bound as in~\cite{Wallman_2014} (instead of directly plugging in the perturbed sequence fidelities), is to perturb each term from the average, in the original unperturbed variance $\mbb{V}[\mc{Z}_m]:=\mbb{E}[\mc{Z}_m^2]-\mc{F}_m^2$, as
\begin{align}
    \mbb{V}\left[\mc{Z}_m^{(g)}\right] &= \mbb{E}\left[\mc{Z}_m^2\right] + \delta\mbb{E}\left[\mc{Z}_m^2\right]- \mc{F}_m^2 - \delta\left(\mc{F}_m^2\right) \nonumber\\
    &= \mbb{E}\left[\mc{Z}_m^2\right] + \delta\mbb{E}\left[\mc{Z}_m^2\right] - \mbb{E}[\mc{Z}_m]^2 - \delta\left(\mbb{E}[\mc{Z}_m]^2\right) \nonumber\\
    &= \mbb{V}[\mc{Z}_m] + \delta\mbb{E}\left[\mc{Z}_m^2\right] - \delta\left(\mbb{E}[\mc{Z}_m]^2\right) \nonumber\\
    &= \mbb{V}[\mc{Z}_m] + \mbb{E}\left[\delta\left(\mc{Z}_m^2\right)\right] - \delta\left(\mbb{E}[\mc{Z}_m]^2\right)
\end{align}
where the second line just substitutes the definition of $\mc{F}_m$, the third line identifies the original unperturbed variance, and the last line substitutes $\delta\mbb{E}\left[\mc{Z}_m^2\right]=\mbb{E}\left[\delta\left(\mc{Z}_m^2\right)\right]$; now use
\begin{align}
    |\delta(X^2)| = |(X_0+\delta{X})^2-X_0^2|=|2X_0+\delta(X)||\delta(X)|\leq 2|\delta(X)|,
\end{align}
as in Eq.(102) of~\cite{Wallman_2014}, for some random variable $X$, where $X_0$ stands for the average and $\delta{X}$ the corresponding perturbation from $X_0$, and where the last inequality follows as $X_0+\delta{X}\in[0,1]$, so that
\begin{align}
    \mbb{V}\left[\mc{Z}_m^{(g)}\right] &\leq \mbb{V}[\mc{Z}_m] + \mbb{E}\left|\delta\left(\mc{Z}_m^2\right)\right| + \left|\delta\left(\mbb{E}[\mc{Z}_m]^2\right)\right| \nonumber\\
    &\leq \mbb{V}[\mc{Z}_m] + 2 \mbb{E}|\delta\mc{Z}_m| + 2 |\delta(\mbb{E}[\mc{Z}_m])| \nonumber\\
    &\leq \mbb{V}[\mc{Z}_m] + 4\mbb{E}|\delta\mc{Z}_m|.
\end{align}

Now, $\delta\mc{Z}_m := \left|\sum_{n=1}^{m+1}\epsilon^n \llangle{M}|\hat{\Xi}_m^{(n)}|\,\rho\rrangle\right|$, and the exact same bound in Eq.~\eqref{eq: bound perturb} applies, since the average is superfluous (as discussed above, the relevant terms are in the noise, which generally give rise to $\dimE\dimS$ factors). Thus, here we have $|\mbb{V}\left[\mc{Z}_m^{(g)}\right]-\mbb{V}[\mc{Z}_m]|\leq\delta_{\mc{V}}$ whenever
\begin{align}
    \epsilon \leq \f{\log\left[1 + \delta_{\mc{V}}/\sqrt{16\dimE\dimS}\right]}{(m+1)\sqrt{\dimE\dimS}},
\end{align}
so again using $\log(1+x) \leq x$, for $x>-1$, in the main text, we get an expression differing from $1/4$ for the bound for the \gls{asf}.
\end{document}